\documentclass[fleqn,usenatbib]{aa}  

\usepackage{graphicx,epstopdf}
\usepackage{txfonts}
\usepackage{hyperref}
\usepackage{amsmath}	
\usepackage{amssymb}	
\usepackage{siunitx} 
\usepackage{xcolor}
\usepackage{subfig}

\begin{document}

   \title{Ionised Gas Kinematics in MaNGA AGN\thanks{Tables 1 and 2 are ony available in electronic form at the CDS via anonymus ftp to cdsarc.u-strasbg.fr (130.79.128.5) or via http://cdsweb.u-strasbg.fr/cgi-bin/qcat?J/A+A/}}
   \subtitle{Extents of the Narrow Line and Kinematically Disturbed regions}
   \author{Alice Deconto-Machado
          \inst{1,2,3}, 
          Rogemar A. Riffel\inst{2,3}, 
          Sandro B. Rembold\inst{2,3},
          Gabriele S. Ilha\inst{2,3},
          Thaisa Storchi-Bergmann\inst{4,3},
          Rogerio Riffel\inst{4,3},
          Jaderson S. Schimoia\inst{2,3},
          Donald P. Schneider\inst{5,6},
          Dmitry Bizyaev\inst{7,8},
           Shuai Feng\inst{9,10},
           Dominika Wylezalek\inst{11},
           Luiz N. da Costa\inst{3,12},
           Jana\'ina C. do Nascimento\inst{3,13}
           \and
           Marcio A. G. Maia\inst{3,12}
         }
  \institute{$^1$Instituto de Astrofis\'ica de Andalucía, IAA-CSIC, Glorieta de la Astronomia s/n, 18008 Granada, Spain\\
  $^2$Departamento de F\'isica, CCNE, Universidade Federal de Santa Maria, 97105-900, Santa Maria, RS, Brazil\\ 
$^3$Laborat\'orio Interinstitucional de e-Astronomia - LIneA, Rua Gal. Jos\'e Cristino 77, Rio de Janeiro, RJ - 20921-400, Brazil\\ 
$^4$Departamento de Astronomia, Universidade Federal do Rio Grande do Sul, IF, CP 15051, Porto Alegre 91501-970, RS, Brazil\\ 
$^5$Department of Astronomy and Astrophysics, The Pennsylvania State University, University Park, PA 16802\\ 
$^6$Institute for Gravitation and the Cosmos, The Pennsylvania State University, University Park, PA 16802\\ 
$^7$Apache Point Observatory and New Mexico State University, Sunspot, NM, 88349, USA\\ 
$^8$Sternberg Astronomical Institute, Moscow State University, Moscow, Russia\\ 
  $^9$Key Laboratory for Research in Galaxies and Cosmology, Shanghai Astronomical Observatory, 80 Nandan Road, Shanghai 200030, China\\ 
$^{10}$University of the Chinese Academy of Sciences, No.19A Yuquan Road, Beijing 100049, China\\ 
$^{11}$European Southern Observatory, Karl-Schwarzschildstr. 2, D-85748 Garching bei M\"unchen, Germany\\ 
$^{12}$Observat\'orio Nacional - MCT, Rua General Jos\'e Cristino 77, Rio de Janeiro, RJ - 20921-400, Brazil\\
$^{13}$Universidade do Vale do Para\'iba, Av. Shishima Hifumi, 2911,  12244-000, São Jos\'e dos Campos, SP, Brazil} 
 
  \authorrunning{Deconto-Machado et al.}
   \date{Received February 19, 2021; accepted January 03, 2022}

 
  \abstract
    {Feedback from active galactic nuclei (AGN) in general seems to play an important role in the evolution of galaxies, although the impact of AGN winds on their host galaxies is still pending on detailed analysis. }
   {We analyse the kinematics of a sample of 170 AGN host galaxies as compared to those of a matched control sample of non-active galaxies from the MaNGA survey in order to characterise and estimate the extents of the Narrow Line Region (NLR) and of the kinematically disturbed region (KDR) by the AGN.}
   {We define the observed NLR radius ($r_{\rm NLR,o}$) as the farthest distance from the nucleus within which both $[\textsc{O iii}]/$H$\beta$ and $[\textsc{N ii}]/$H$\alpha$ ratios fall in the AGN region of the BPT diagram and the H$\alpha$ equivalent width is required to be larger than 3.0 $\AA$. The extent of the KDR ($r_{\textrm{KDR,o}}$) is  defined as the distance from the nucleus within which the AGN hosts galaxies shows a more disturbed gas kinematics than the control galaxies.}
   {The AGN $[\textsc{O iii}]\lambda 5007$ luminosity ranges from $10^{39}$ to $10^{41}$ erg s$^{-1}$, and the kinematics derived from the $[\textsc{O iii}]$ line profiles reveal that, on average, the most luminous AGN ($L[\textsc{O iii}] > 3.8 \times 10^{40}$ erg s$^{-1}$) possess higher residual difference between the gaseous and stellar velocities and velocities dispersion than their control galaxies in all the radial bins. Spatially resolved NLR's and KDR's were found in 55 and 46 AGN host galaxies,
   with corrected radii $0.2< r_{\rm KDR,c}< 2.3$ kpc  and   $0.4< r_{\rm NLR,c} < 10.1$ kpc, with a relation between the two given by $\log r_{\rm KDR,c} = (0.53\pm 0.12)\,\log r_{\rm NLR,c} +(1.07\pm 0.22)$, respectively. On average, the extension of the KDR corresponds to about 30 per cent of that of the NLR. Assuming that the KDR is due to an AGN outflow, we have estimated ionised gas mass outflow rates that range between $10^{-5}$ and $\sim 1$\,M$_{\odot}$\,yr$^{-1}$, and kinetic powers that range from $10^{34}$ to $10^{40}$\,erg\,s$^{-1}$.}
   {Comparing the power of the AGN ionised outflows with the AGN luminosities, they are always below the 0.05 L$_{\textrm{AGN}}$ model threshold for having a an important feedback effect on their respective host galaxies. The mass outflow rates (and power) of our AGN sample correlate with their luminosities, populating the lowest AGN luminosity range of the correlations previously found for more powerful sources.}
   

   \maketitle
%

\section{Introduction}

 \par Only a small fraction ($\sim$10\,\%) of galaxies are expected to present an active galactic nucleus \citep[AGN,][]{Ho2008}. This phenomenon occurs when the supermassive black hole (SMBH) in the centre of the galaxy captures nearby matter. 
During this process, this matter is gradually accreted, turning gravitational potential energy into electromagnetic radiation and kinetic energy, by ejection of particles in the form of winds.  The radiation and winds produced in the accretion disk could play an important role in the evolution of the AGN host galaxy -- the so-called AGN feedback mechanisms \citep{cattaneo09,fabian12,harrison17}. This feedback may influence the evolution of the galaxy, as predicted in cosmological simulations \citep{croton}.  

\par In the unified scheme of AGN \citep{antonucci, urry}, a torus surrounding the accretion disk collimates its radiation leading to a bi-conical morphology for the Narrow-Line Region (NLR). A similar structure could be expected for the winds launched from the accretion disc which could also be collimated by the torus. Narrow-band HST images of type 2 QSOs reveal elongated ENLRs and bipolar ionisation cones whose extents increase with $[\textsc{O iii}]\lambda5007$ luminosity \citep{fischer18,thaisa2018}, but in less-luminous Seyfert galaxies the conical morphology is not as common as expected from the unified scheme \citep{schmitt2003}.  \citet{He_2018} found bipolar structures with an opening angle of $\sim 80^{\circ}$ in NLRs from MaNGA AGN by analysing emission lines ratio behaviours.

The $[\textsc{Oiii}]\lambda 5007$ kinematics of the inner kilo-parsec of nearby AGN reveal that although bi-conical outflows are observed, they appear to be more the exception than the rule. For instance, \citet{mullaney} find that AGN driven winds due to the interaction of radio jets with the ambient gas usually present a linear structure and not a bi-conical morphology. \citet{fischer2013} has also studied the  $[\textsc{Oiii}]\lambda 5007$ kinematics in the NLR of 48 Seyfert galaxies and found that in only 12 objects the ionised gas kinematics is consistent with conical outflows, indicating that the gas kinematics of AGN is more complex than predicted by the unified model. \citet{fischer18} investigated the $[\textsc{Oiii}]\lambda 5007$ emission distribution and kinematics in a sample of 12 type 2 nearby ($z < 0.12$) and luminous (log $L_{\rm [OIII]}/{\rm erg s^{-1}} \gtrsim 42$ quasars, using HST observations. They found that the gas is kinematically disturbed by the AGN up to distances of $\sim$ 8-15 kpc and determined  
an average maximum outflow radius of $\sim$600 pc.
\cite{wylezalek2020} discussed the ionised gas kinematics of a sample of AGN observed with MaNGA and found that high velocity gas is more prevalent in AGN compared to non-AGN, based on measurements of the line velocity width that encloses 80\% of the total flux of the $[\textsc{Oiii}]\lambda 5007$ emission line.

\par AGN-driven outflows are known to extend from tens to kpc scales, achieving velocities of hundreds of km s$^{-1}$ \citep{crenshaw2003,ilha2019,wylezalek2020}. 
However, the impact of these outflows in the AGN host galaxies remains unclear, and recent studies have focused in this question by studying the gas kinematics on large (kpc) scales.  Using spatially resolved long-slit spectroscopy, \citet{greene2011} investigated the kinematics of a sample of 15 nearby ($z < 0.5$) quasars with [\ion{O}{III}]$\lambda$5007 luminosity $>10^{42}$ erg s$^{-1}$ and reported that the AGN is the main ionising agent of the interstellar medium over the entire structure of their host galaxies.
In powerful AGN, large-scale outflows are detected, and their extents increase with the AGN luminosity \citep{greene2012, fischer18,thaisa2018}. By analysing the $[\textsc{Oiii}]\lambda 5007$ and H$\beta$ emitting gas kinematics of a sample of 14 quasars, observed with Gemini IFUs, \citet{liu2013} found that all objects possess extended ionised gas nebulae outflows with mean diameter of $28$ kpc. Such powerful outflows are able to halt star formation in the host galaxy \citep[e.g.][]{alatalo15}. However, \citet{husemann} find that the results from \citet{liu2013} and also \citet{liu2014} may be affected by not considering beam-smearing effects which indicates that probably the outflows present smaller extents \citep{villarmartin,Karouzos_2016, tadhunter}.

\par The role of the AGN feedback in low luminosity AGN can be addressed with the current surveys of integral field spectroscopy of galaxies, such as the Sloan Digital Sky Survey - IV's \citep[SDSS-IV][]{blanton} Mapping Nearby Galaxies at APO \citep[MaNGA; ][]{bundy2015} and Calar Alto Legacy Integral Field Area \citep[CALIFA; ][]{sanchez12}.  We have been using MaNGA data to investigate the AGN hosts properties, with results presented in a series of papers, as follows. \citet{rembold2017} discusses nuclear stellar population properties of the first 62 AGN observed with the MaNGA survey and defined a control sample of inactive galaxies that match the AGN hosts properties in terms of stellar mass, redshift, visual morphology and inclination. It was found that AGN of increasing luminosity exhibit an increasing contribution from the youngest stellar population relative to control galaxies and a decrease in the oldest components. 

\par The sample defined by \citet{rembold2017} was used in subsequent studies.  \citet{mallmann2018} presented spatially resolved stellar population properties using the MaNGA data. They have shown that the fraction of young stellar population in high-luminosity AGN is higher in the inner ($R\le 0.5 R_e$, where $R_e$ is the galaxy effective radius) regions when compared with the control sample. The low-luminosity AGN and control galaxies display similar fractions of young stars over the whole MaNGA field-of-view ($\sim1\,R_e$). \citet{nascimento2019} focused on the emission-line flux distributions and gas excitation of this sample and reported that the extent of the region ionised by the AGN is proportional to $L_{[\textsc{Oiii}]}^{0.5}$, where $L_{[\textsc{Oiii}]}$ is the luminosity of the  $[\textsc{Oiii}]\lambda 5007$ emission line. They also found that the star formation rate is higher in AGN than in the control galaxies for the early-type host galaxies. The analysis of the nuclear stellar and gas kinematics, as well as the difference of the orientation of the line of nodes derived from the stellar and gas velocity fields (kinematic position angle [PA] offset), is presented in \citet{ilha2019}. By comparing AGN host and inactive galaxies, no difference was found in terms of the kinematic PA offsets between gas and stars. But AGN present higher gas velocity dispersion within the inner 2\farcs5 diameter region, that has been interpreted as due to AGN-driven outflows.

\par In the present study, we use an expanded sample relative to that of the above papers, consisting of all AGN included in the MaNGA Product Launch 8 (MPL-8) \citep{law2016}, to investigate the large scale gas kinematics of AGN hosts, and compare them to those of a control sample of inactive galaxies, drawn in the same way as in \citet{rembold2017}. Our AGN sample consists of 170 AGN and two control galaxies for each AGN host. 

\par This paper is organised as follows: Section \ref{sec:data} presents a brief description of the MaNGA survey and discusses the sample properties. In Sec. \ref{measurements} we describe our measurements. In Sec.~\ref{results}, we present radial profiles of the kinematic properties of the sample galaxies, which are discussed in Sec.~\ref{disc}. The main conclusions of this work are summarised in Sec.~\ref{conc}. The assumed cosmological parameters in this work are $H_0=70$ km s$^{-1}$ Mpc$^{-1}$, $\Omega_m=0.3$, and $\Omega_V=0.7$.

\section{Data}\label{sec:data}
\par The MaNGA survey's focus is on understanding the formation and evolution of galaxies by mapping the kinematics and physical properties of $\sim10\,000$ nearby galaxies, with redshift ($z$) ranging from 0.01 to 0.15 and stellar masses greater than $10^9$\,M$_\odot$ \citep{law2015}. The integral field unit (IFU) spectroscopic observations have been done with the 2.5-m Sloan telescope of the Apache Point Observatory (APO) \citep{gunn, smee, wake, yan2016a, yan20162}. The  spectral resolving power is $R \sim 2000$, resulting in an instrumental broadening of $\sigma_{\rm inst}=70$ km\,s$^{-1}$ \citep{law2021}, and the average IFU's angular resolution is $2\farcs5$ \citep{yan20162,law2016}. Details about the survey, instrument, and data processing are given in \citet{bundy2015} and \citet{drory2015}.  

\par SDSS-IV Data Release (DR) 14 \citep{abolfathi2018} includes 2778 MaNGA data cubes.  \citet{rembold2017} presented a sample of all AGN observed by MaNGA in DR14, as obtained from diagnostic diagrams  \citep{bpt,whan}, hereafter called BPT and WHAN diagrams, based on emission-line fluxes and equivalent widths measured using the SDSS-III spectra by \citet{thomas13}. These diagrams can be used to identify the dominant gas ionisation mechanism in each galaxy and are among the most commonly used diagnostic diagrams to select AGN using optical observations. To be selected as an AGN host, the galaxy must be located in the Seyfert or LINER region on both BPT and WHAN diagrams simultaneously, including the uncertainties. 
 In order to analyse the effects of AGN on the host galaxies, \citet{rembold2017} have also selected non-active galaxies to compose a control sample. Two control galaxies were selected for each AGN. This selection was performed by matching the morphology, stellar mass, absolute magnitude and redshifts of the inactive galaxies to those of the AGN hosts. The initial control sample contains 124  galaxies (twelve of them were paired to more than one AGN host). The initial AGN and control samples, as well as their main properties, are listed in \citet{rembold2017}.

\par
\begin{figure*}
    \centering
    \subfloat[]{\label{bptw1}\includegraphics[width=0.5\linewidth]{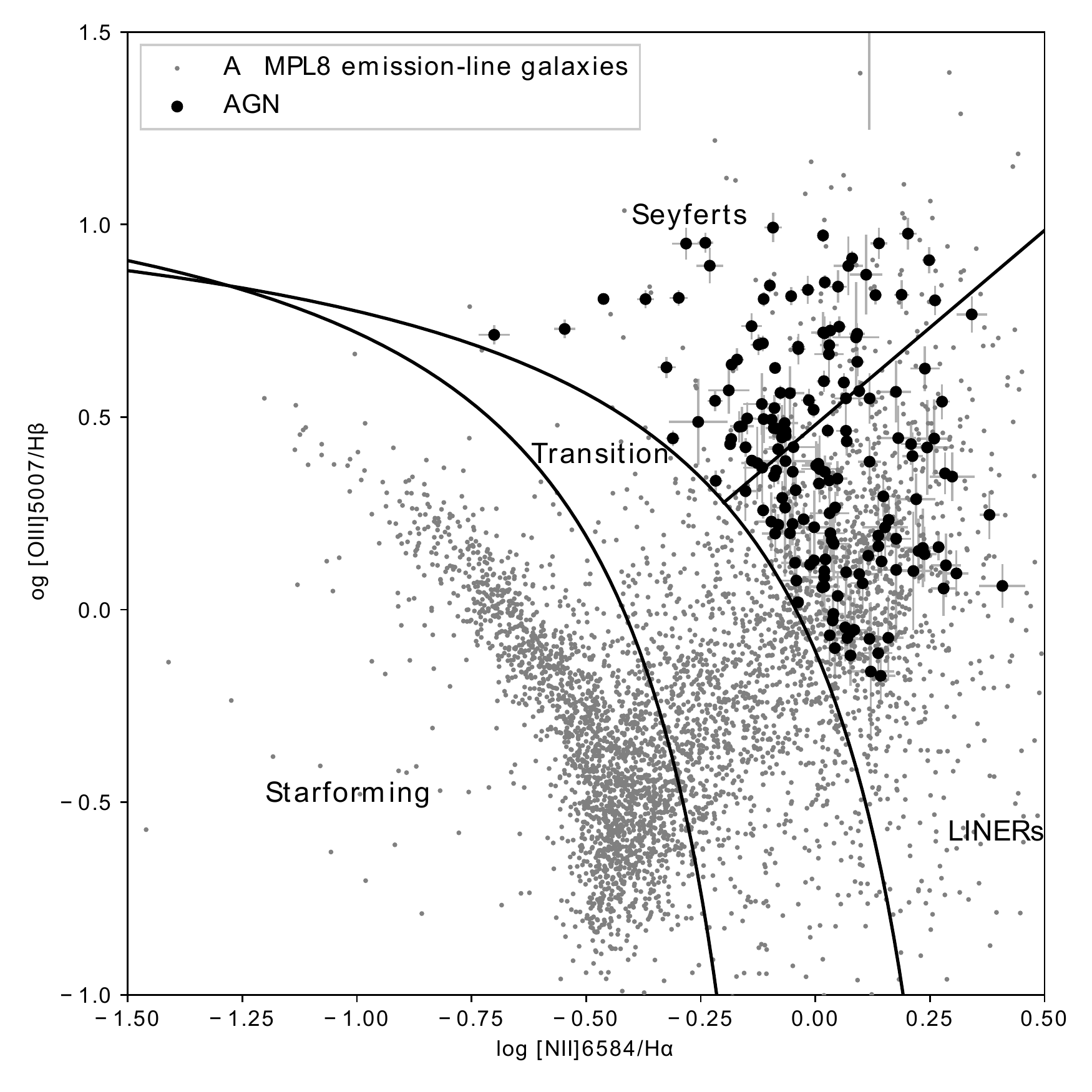}}
\subfloat[]{\label{bptw2}\includegraphics[width=0.5\linewidth]{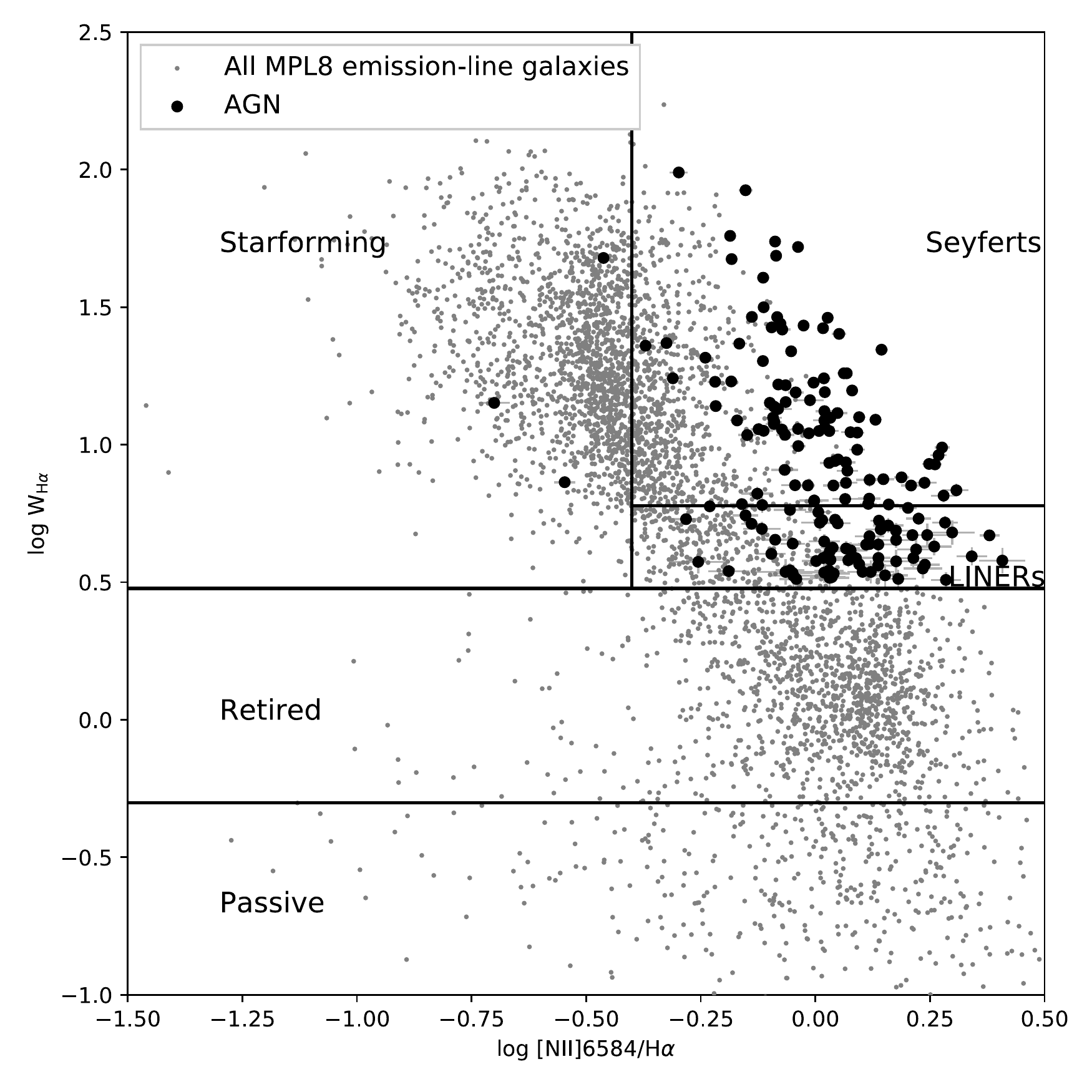}}\\
\subfloat[]{\label{bptw3}\includegraphics[width=0.5\linewidth]{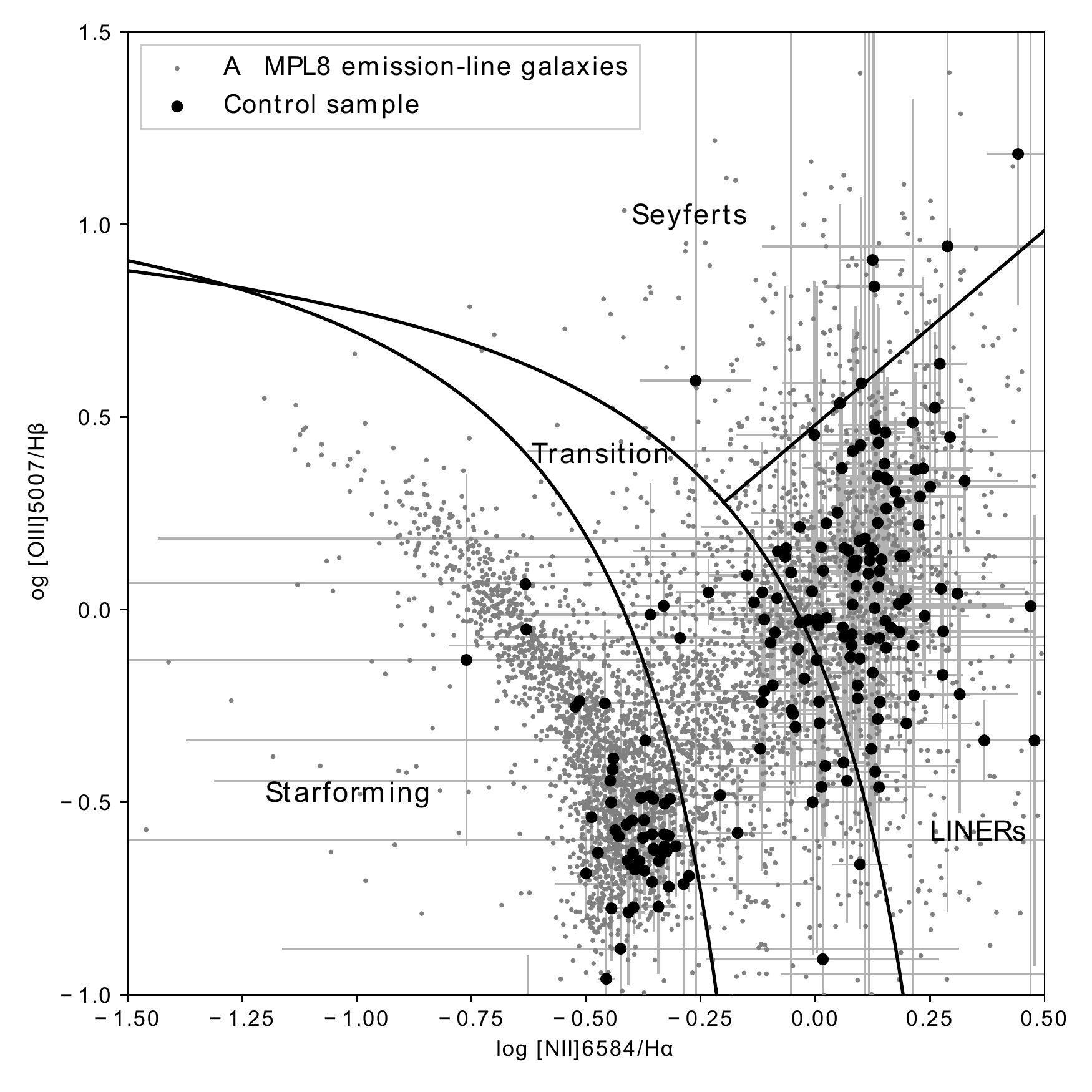}}
\subfloat[]{\label{bptw4}\includegraphics[width=0.5\linewidth]{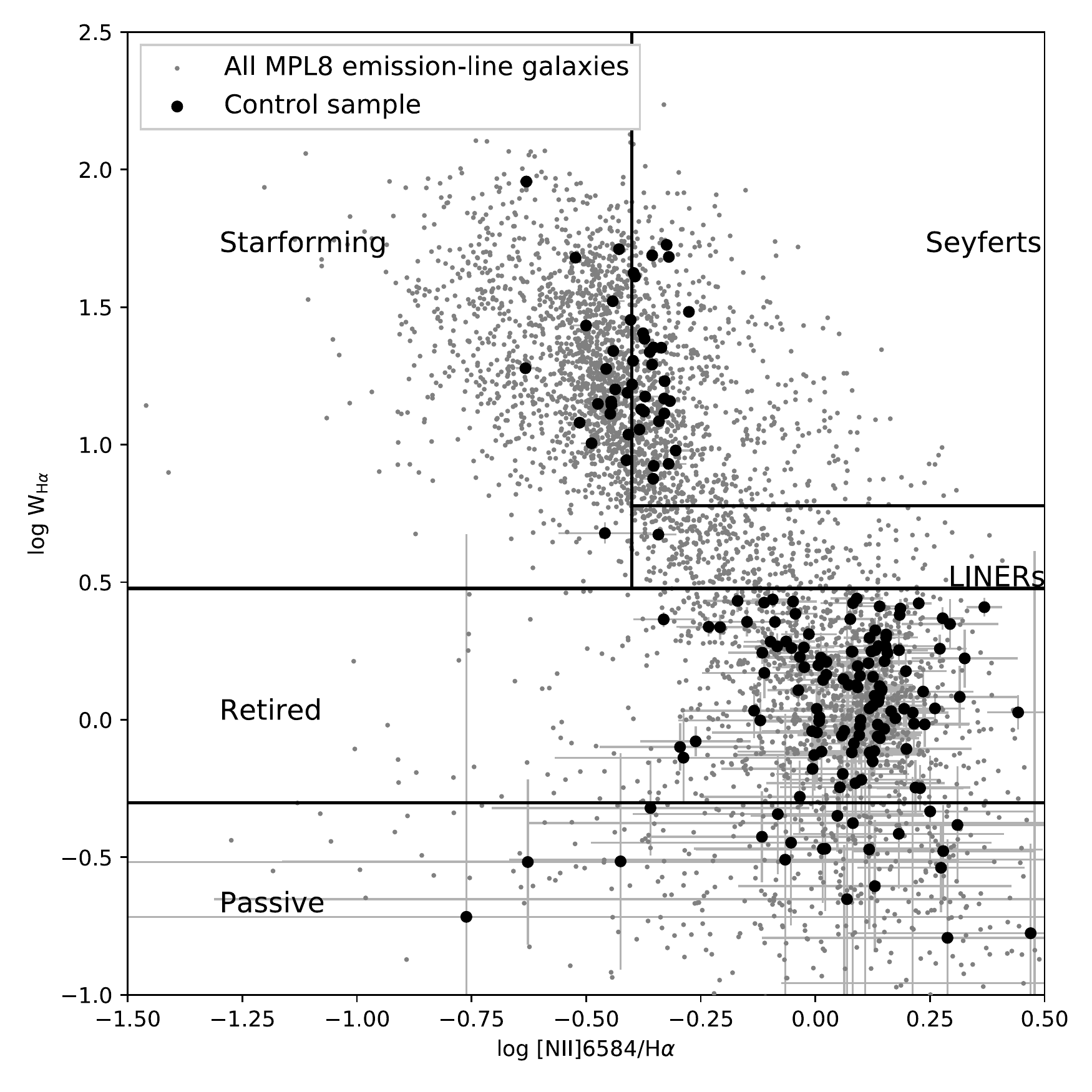}}
\caption{BPT (left) and WHAN (right) diagrams for the AGN hosts (top) and control objects (bottom) in our sample. The separating lines in the plots are from \citet{kauffmann2003}, \citet{kewley01} and \citet{whan}. Grey dots represent all the emission-line galaxies in MPL-8. Large black circles represent AGN and control objects.}
    \label{fig:diagnostic}
\end{figure*}

\begin{figure*}
    \centering
    \includegraphics[width=\textwidth]{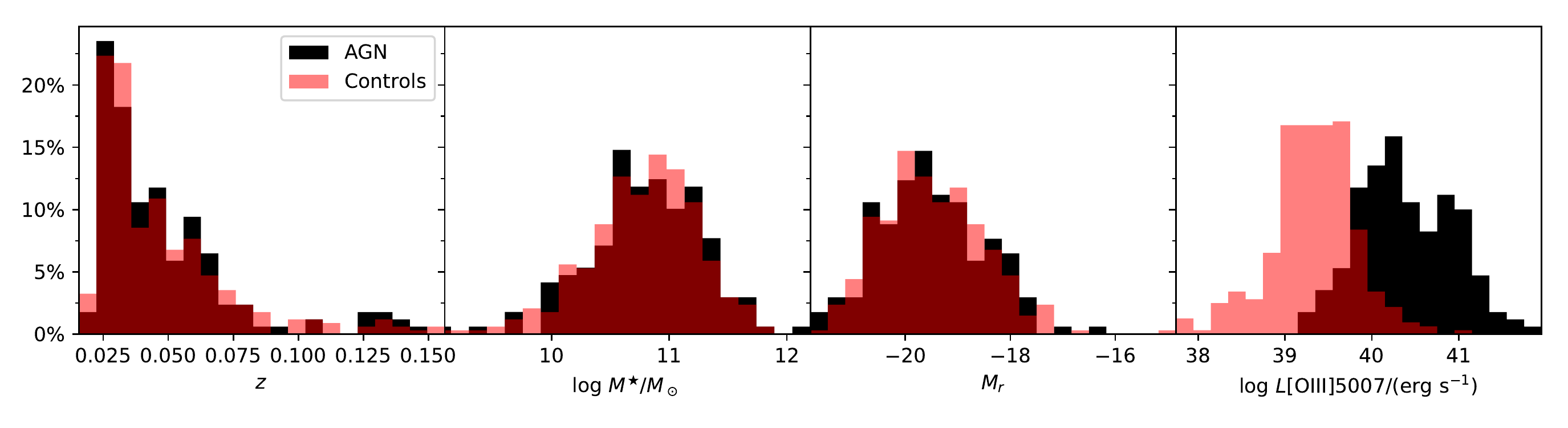}
    \caption{Redshift ($z$), stellar mass ($M^\bigstar$), integrated \textit{r}-band ($M_r$) absolute magnitude and $[\textsc{O iii}]\lambda 5007$ distributions for AGN (black) and control samples (red). Both samples are composed mostly of low-luminosity galaxies.}
    \label{fig:hist_params}
\end{figure*}

\begin{figure*}
    \centering
    \begin{tabular}{ccc}
AGN & Control object 1 & Control object 2 \\
1-174945 & 1-137845 & 1-135372\\
\includegraphics[height=3.5cm]{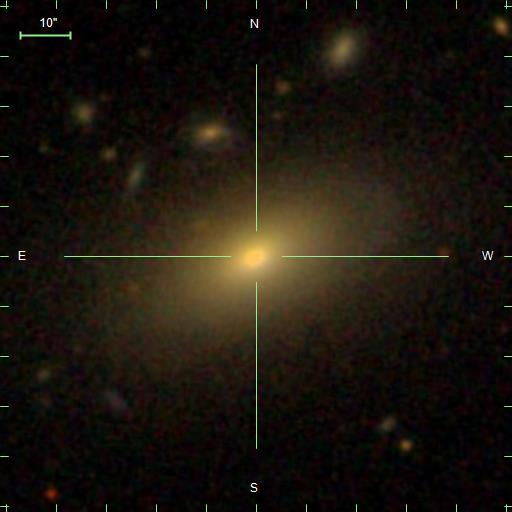} & \includegraphics[height=3.5cm]{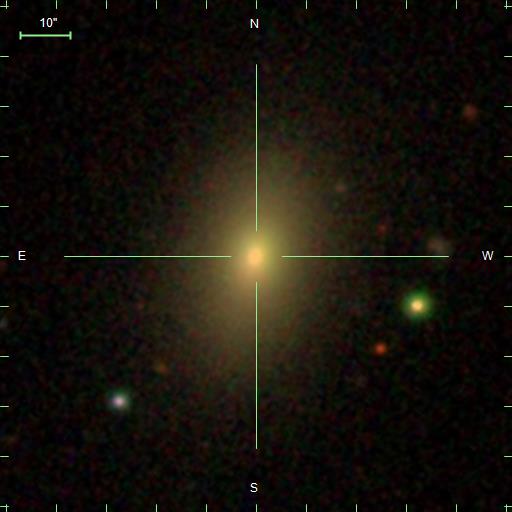} & \includegraphics[height=3.5cm]{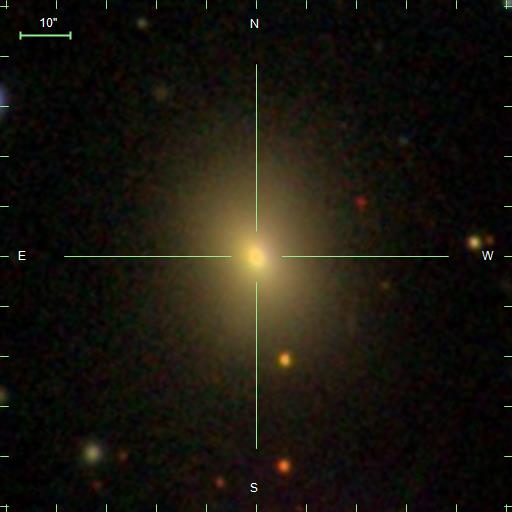} \\
1-297172 & 1-456656 & 1-395508\\
\includegraphics[height=3.5cm]{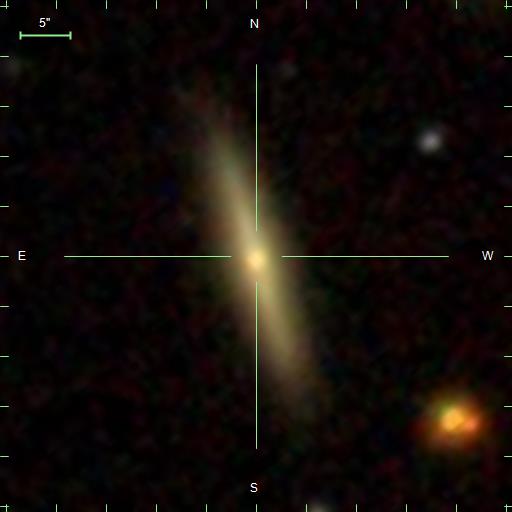} & \includegraphics[height=3.5cm]{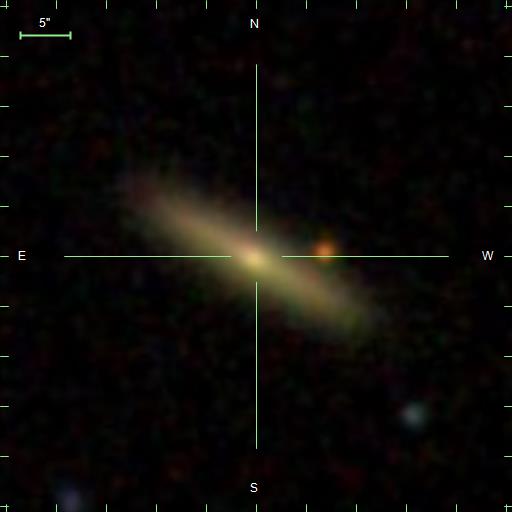} & \includegraphics[height=3.5cm]{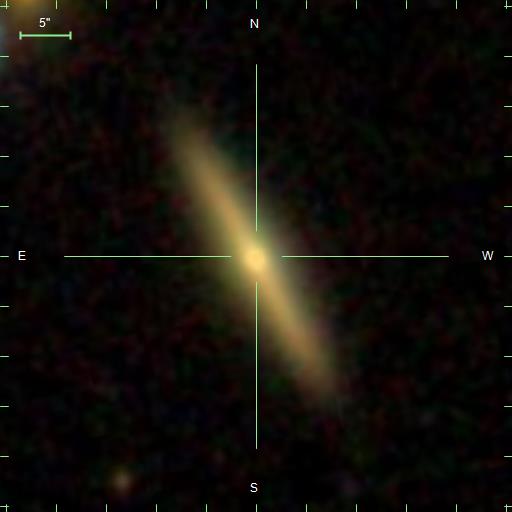} \\
1-31788 & 1-379708 & 1-317822\\
\includegraphics[height=3.5cm]{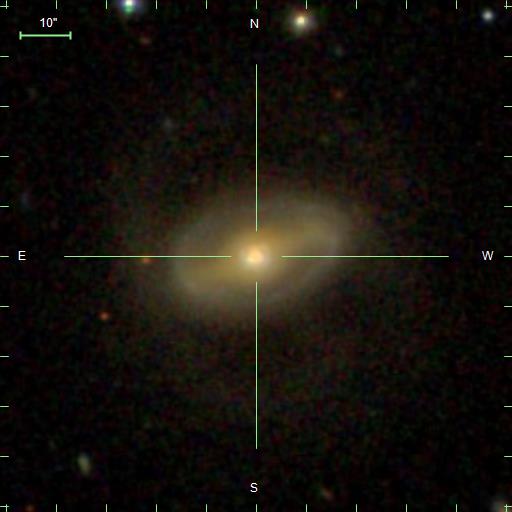} & \includegraphics[height=3.5cm]{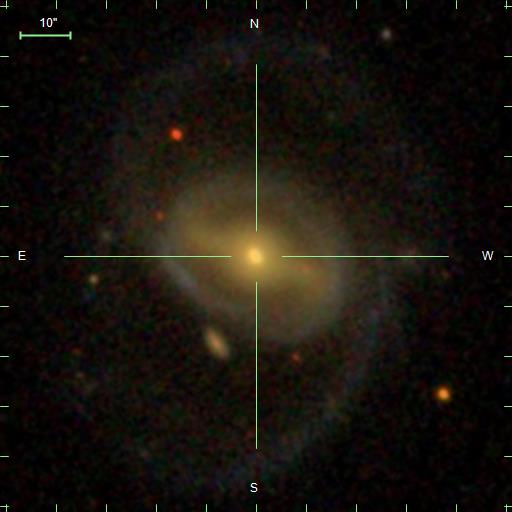} & \includegraphics[height=3.5cm]{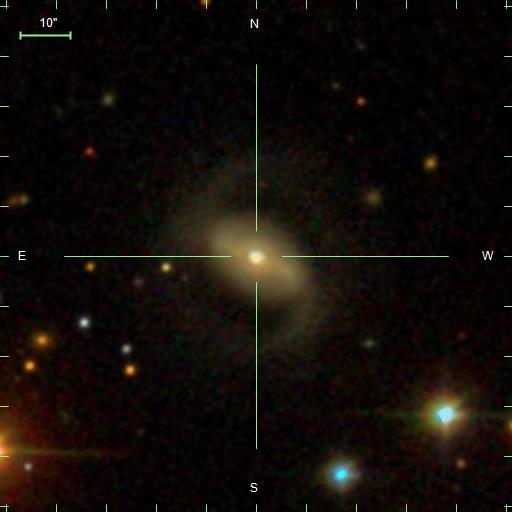} \\
1-232143 & 1-173966 & 1-627397\\
\includegraphics[height=3.5cm]{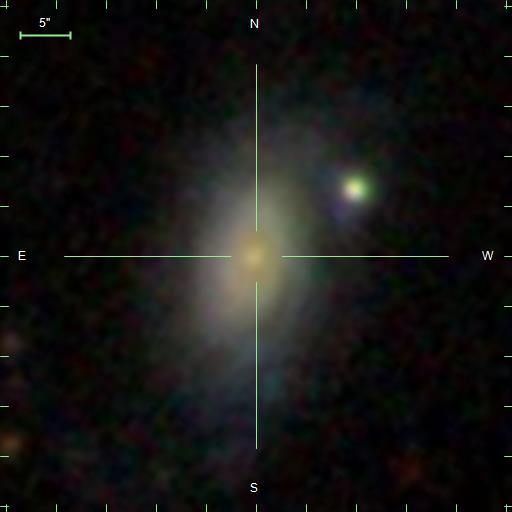} & \includegraphics[height=3.5cm]{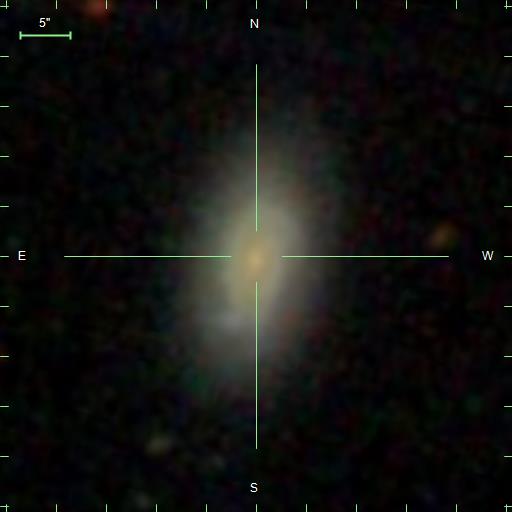} & \includegraphics[height=3.5cm]{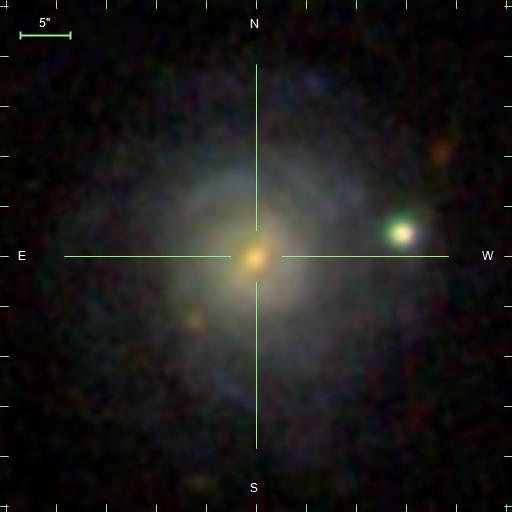} \\
\end{tabular}
        \caption{SDSS-III multicolor images of four representative examples of AGN hosts (left column) and their respective control galaxies (centre and right columns) in our sample. These galaxies were chosen to roughly reproduce the morphological Hubble sequence from ellipticals (top) to late-type spirals (bottom).}
    \label{fig:examples_agn_cs}
\end{figure*}

 After the release of the MaNGA Product Launch 8 (MPL-8), which contains data-cubes for 6779 galaxies, the number of observed AGN with MaNGA has grown to 173 objects  using the same criteria as in \citet{rembold2017}, including the 62 AGN presented in that work. Following \citet{rembold2017}, we selected two control galaxies for each new AGN except for three objects: the galaxies identified as MaNGA ID 1-37440, 1-189584, and 31-115. Since these galaxies have low redshifts ( $z\lesssim$0.01), which are close to the lower limit of MaNGA's redshift range, we were not able to identify control galaxies at similar redshifts. Considering this issue and the fact that the main goal of this paper is to compare the gas kinematics of AGN hosts and their control galaxies, we do not include these objects in the global statistics. The final AGN sample consists of 170 galaxies, with a $[\textsc{O iii}]\lambda 5007$ emission line luminosity ranging from $10^{39}$ to $10^{41}$ erg s$^{-1}$, as measured from the SDSS-III spectra. The AGN and their respective control galaxies are listed in Tables 1 and 2 (available in electronic form only), respectively. Fig.~\ref{fig:diagnostic} presents the location of AGN hosts and control galaxies of our sample in the diagnostic diagrams BPT and WHAN. Fig. \ref{fig:hist_params} shows the distributions of redshift ($z$), stellar mass ($M^\bigstar$), \textit{r}-band ($M_r$) absolute magnitude and $[\textsc{O iii}]\lambda 5007$ luminosity ($L[\textsc{Oiii}]$) of our AGN sample, as compared to that of their associated control galaxies. As in the initial sample \citep{rembold2017}, the AGN and control samples present similar distributions in all parameters except ($L[\textsc{Oiii}]$), which is shifted towards larger
values for the AGN hosts. Fig.~\ref{fig:examples_agn_cs} presents the SDSS-III multicolor images of four representative AGN hosts in our extended sample -- i.e. those not included in \citet{rembold2017} -- and their associated controls.

\par In total, the AGN sample contains 57 early-type (33.5$\%$), 87 late-type (51.2$\%$), and 3 merger galaxies (1.8$\%$), according to the Galaxy Zoo database. There is no classification in Galaxy Zoo for 23 objects (13.5$\%$). Similarly, the control sample is composed of 125 early-type (36.8$\%$), 182 late-type (53.5$\%$), 4 merger galaxies (1.2$\%$), and 29 objects (8.5$\%$) whose classifications are uncertain. Regarding the AGN classification, 94 (55.3\%) out of the 170 sources present LINER nuclei, while the other 76 (44.7\%) galaxies present Seyfert nuclei, as indicated by the BPT diagram (Fig. \ref{fig:diagnostic}).  By visual inspection of the nuclear spectra of all AGN hosts, we conclude that 12 (7.06 $\%$ of the sample) objects are type 1 AGN (showing broad components in their hydrogen recombination lines). The fraction of type 1 AGN in our sample is smaller than that obtained using SDSS-DR7 spectra, of 10--34 $\%$ for similar [O\,{\sc iii}] luminosities \citep{oh15}. This difference is likely due to selection effects of the MaNGA sample and by the visual inspection of the observed spectra, which may not be sensitive to faint broad line components.  These weak broad components do not affect the determinations of NLR properties, which is the aim of this study.

We have also separated the sample according to the $[\textsc{O iii}]\lambda 5007$ emission line luminosity (L$[\textsc{O iii}]$), which can be used as a proxy for the bolometric luminosity of the AGN \citep{heckman_2004}. The $[\textsc{O iii}]\lambda 5007$ luminosity L$[\textsc{O iii}]$ is determined within a circular aperture of 2\farcs5 diameter from the nucleus, based in a single fiber.  Following previous works by our group \citep{rembold2017,mallmann2018,ilha2019,nascimento2019} we separate our sample into strong and weak AGN: strong AGN are defined as those with $L[\textsc{O\,iii}] \ge 3.8 \times 10^{40}$ erg s$^{-1}$, measured within a nuclear aperture of 3$''$ diameter; weak AGN are those with $L[\textsc{Oiii}] < 3.8 \times 10^{40}$ erg s$^{-1}$. We use the same $L[\textsc{Oiii}]$, originally proposed by \citet{kauffmann2003} and that roughly divide their sample in Seyfert (defined by the authors as objects with [O {\sc iii}]5007/H$\beta>3$  and [N {\sc ii}]6583/H$\alpha>0.6$) and LINER ([O {\sc iii}]5007/H$\beta<3$  and [N {\sc ii}]6583/H$\alpha>0.6$) nuclei.
In the complete sample, 62 galaxies can be classified as strong AGN, which corresponds to 36$\%$ of our AGN sample; the new sample includes almost four times the number of strong AGN when compared to the 17 of \citet{rembold2017}.  The number of strong AGN is slightly smaller than the number of Seyfert nuclei (76) and for consistency with our previous works, we will separate our sample into strong and weak AGN in this study.   The larger number of strong AGN is relevant, as the main differences in terms of physical properties of the gas and stars of AGN and non-AGN hosts in previous works \citep{mallmann2018,ilha2019,nascimento2019} were seen in strong AGN. \citet{wylezalek2020} also observed that the ionised gas kinematics is more extreme in more luminous MaNGA AGN.

The most important difference between our sample and those of previous studies is that our sample is based on MPL-8, which include data for 6779 galaxies, while the previous works were based on the Data Release (DR) 14 \citep{dr14}, including 2744 galaxies. In addition, this work and that of \citet{wylezalek2020} used distinct AGN selection criteria. While our selection is based on SDSS-III spectroscopic data from DR 12 \citep{dr12}, \citet{wylezalek2020} used MaNGA data, including off centred photo-ionisation signatures to classify a galaxy as an AGN \citep{wylezalek18}. Finally,  an advantage of our work is the selection of a control sample of non-active galaxies with properties matching those of the AGN hosts. This allows us to better investigate the effects of AGN feedback on the host galaxy properties, by making sure that eventual differences are not related to properties such as stellar mass or Hubble type.

\section{Measurements}\label{measurements}

\par The measurements of the emission-line fluxes and the gas and stellar kinematics from the MaNGA data cubes were performed  using the Gas AND Absorption Line Fitting (\textsc{gandalf}) code \citep{sarzi2006}, written in Interactive Data Language (IDL). This algorithm maps the gas emission and kinematics while separating the stellar continuum and the gas contributions to the galaxy's spectra. In order to subtract the continuum spectra and also to measure the stellar kinematics, \textsc{gandalf} uses the Penalized Pixel-Fitting (\textsc{pPXF}) code \citep{capellari, capellari2}. The \textsc{pPXF} code consists of stellar absorption lines fitting, using templates spectra as bases. Following \citet{ilha2019},  we used 30 selected Evolutionary Population Synthesis models from
\citet{bruzual-charlot} as template spectra. These models cover ages ranging from 5~Myr to 12~Gyr and three metallicities ($0.004\,Z_{\odot},0.02\,Z_{\odot},0.05\,Z_{\odot}$).  In addition, we included multiplicative Legendre polynomial of order 3 to account for the different continuum shapes and the line of sight velocity distribution is represented by a Gaussian.

\begin{figure*}
\begin{center}
MaNGA 1-48116
\end{center}
\centering
\includegraphics[width=0.95\textwidth]{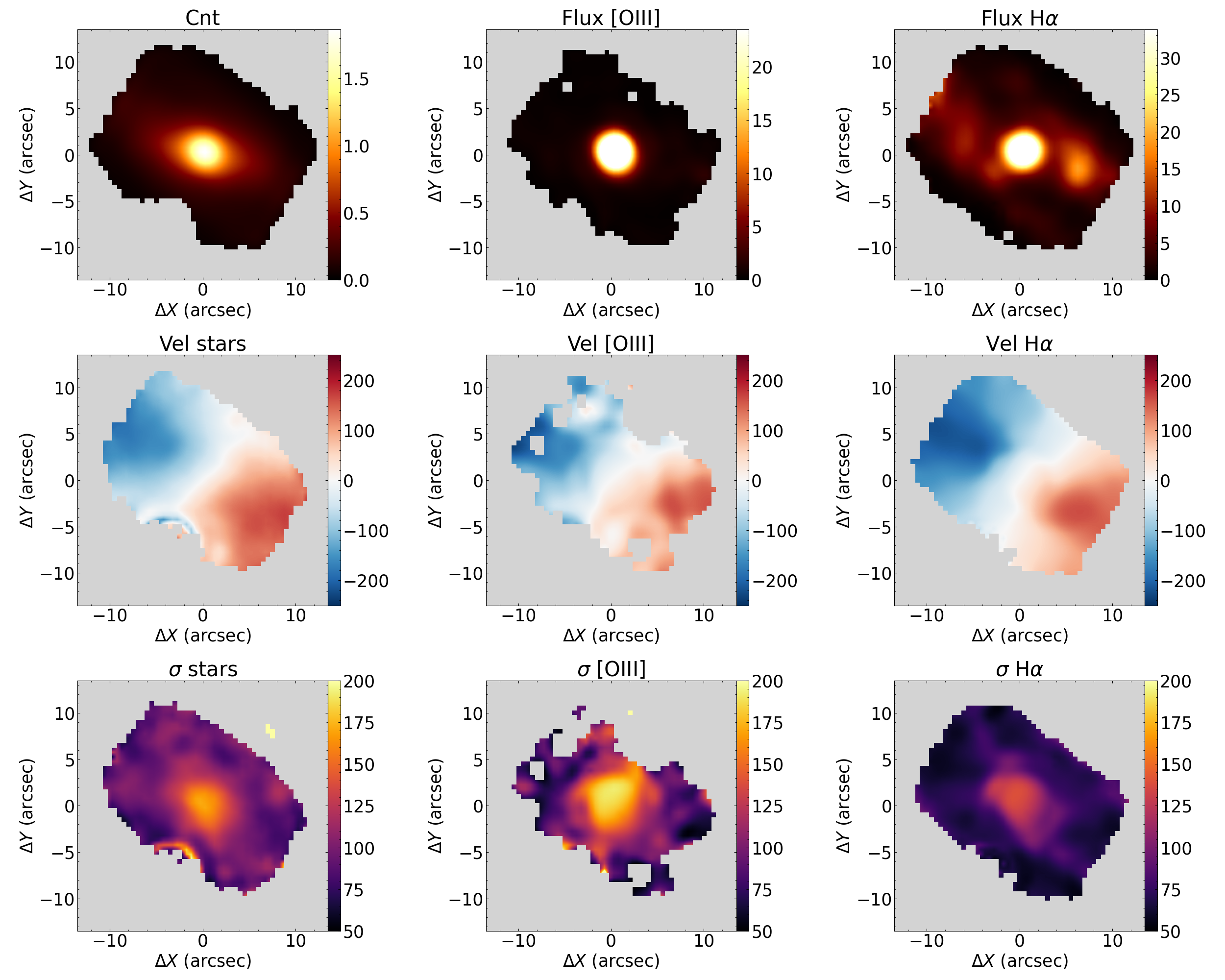}
\caption{Example of maps we have obtained from our measurements for the galaxy MaNGA 1-48116. The first row shows, from left to right, a continuum map in units of $10^{-17}\,{\rm erg s^{-1} cm^{-2} \AA^{-1}}$ obtained by collapsing the whole spectral range, the  [O\,{\sc iii}]5007\,\AA\ and H$\alpha$ flux maps in units of $10^{-17}\,{\rm erg s^{-1} cm^{-2}}$. The second row shows, from left to right, the stellar, [O\,{\sc iii}] and H$_\alpha$ velocity fields in km\,s$^{-1}$, relative to the systemic velocity of the galaxy. The bottom row shows the velocity dispersion maps for the stars, [O\,{\sc iii}], and H$_\alpha$, from left to right. In all panels, the North points up and East to the left and the $\Delta X$ and $\Delta Y$ labels show the distance relative to the peak of the continuum emission.  } 
\label{fig:example}
\end{figure*}

The following emission lines were fitted: H$\beta$, [O\,{\sc iii}]$\lambda\lambda$4959,5007, [O\,{\sc i}]$\lambda$6300, [N\,{\sc ii}]$\lambda\lambda$6548,6583, H$\alpha$ and [S\,{\sc ii}]$\lambda\lambda$6716,6731. Each emission line is fitted by a single Gaussian component, except for type 1 objects, where a broad component is included to reproduce the nuclear profiles of H$\beta$ and H$\alpha$. The centroid velocity and width of the doublets [O\,{\sc iii}]$\lambda\lambda4959,5007$, [N\,{\sc ii}]$\lambda\lambda6548,6583$ and [S\,{\sc ii}]$\lambda\lambda6716,6731$  emission lines were kept tied, for each pair separately. During the fit, the following intensity ratios were kept fixed to their theoretical values: [O\,{\sc iii}]$\lambda5007$/[O\,{\sc iii}]$\lambda4959=2.86$ and [N\,{\sc ii}]$\lambda6583$/[N\,{\sc ii}]$\lambda6548=2.94$ \citep{osterbrock06}. As discussed in \citet{ilha2019}, the emission-line profiles are well reproduced by single-Gaussian fits; details about the spectral fitting can be found there. 
In Figure\,\ref{fig:example} we present an example of the resulting maps of the properties we have derived from our measurements for the AGN host with {\it{mangaid}} 1-48116.

\begin{figure*}
    \centering
    \includegraphics[scale=0.46, trim=0.9cm 0cm 0cm 0cm, clip]{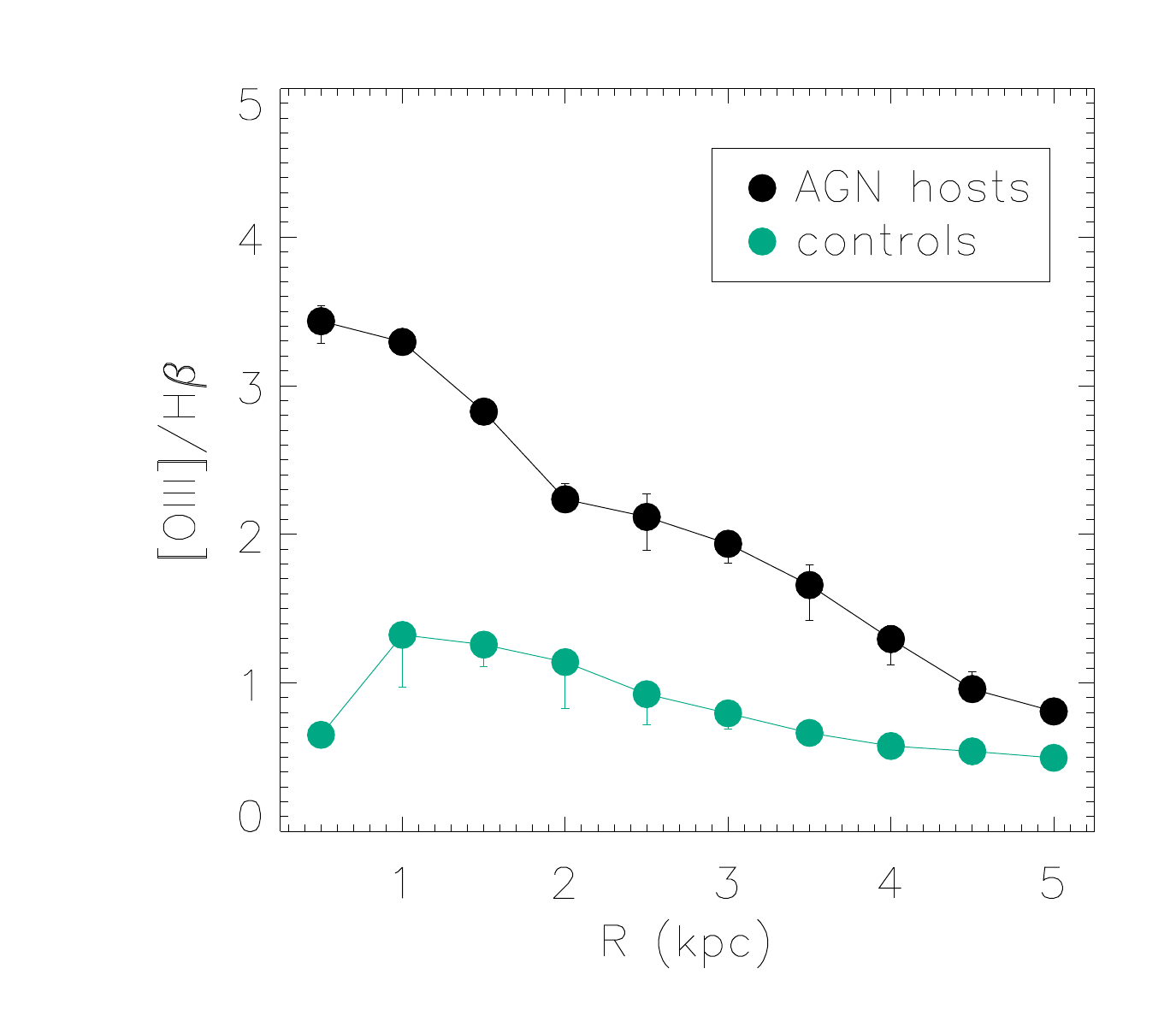}
    \includegraphics[scale=0.46, trim=0.9cm 0cm 0cm 0cm, clip]{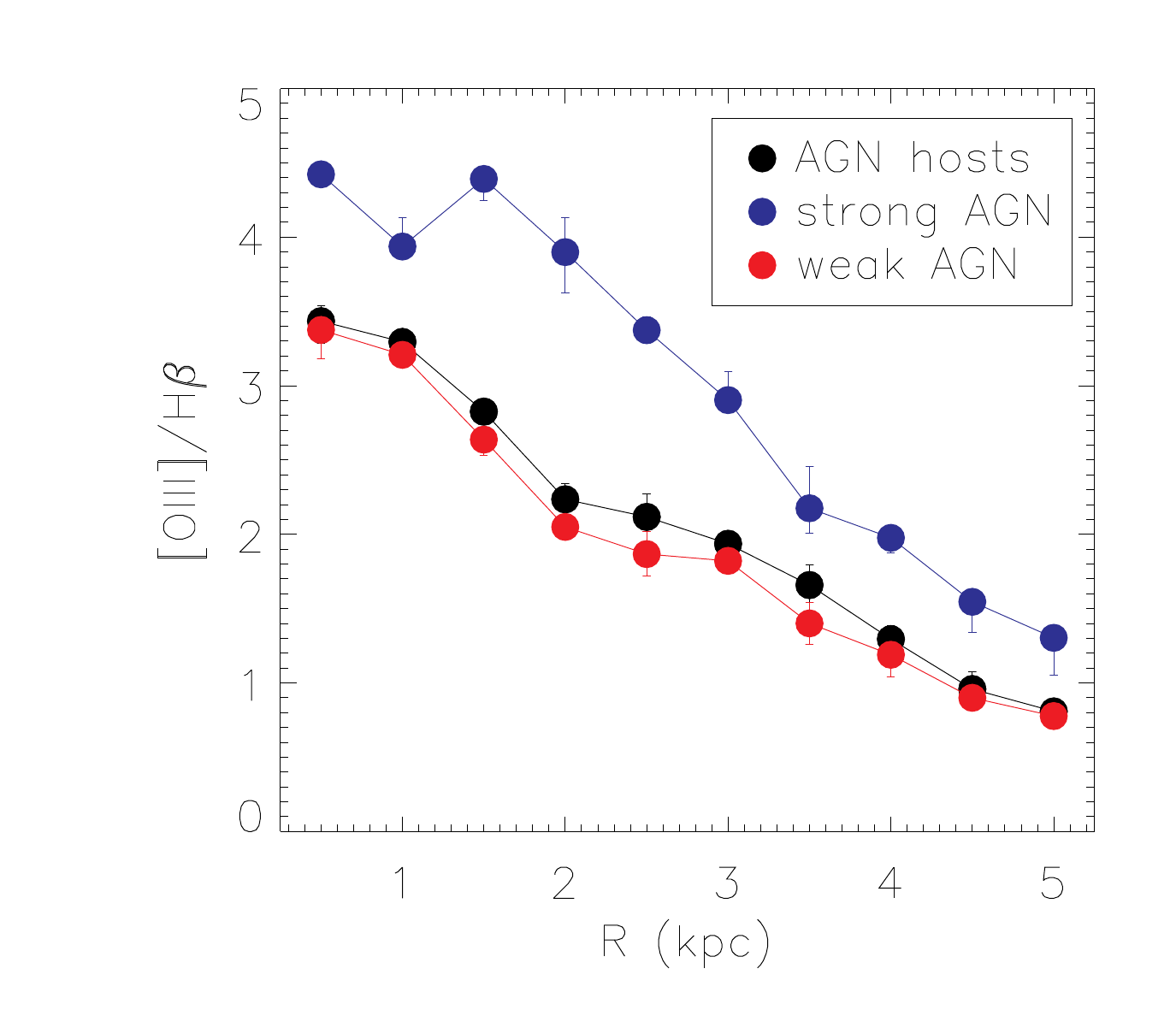}
    \includegraphics[scale=0.46, trim=0.9cm 0cm 0cm 0cm, clip]{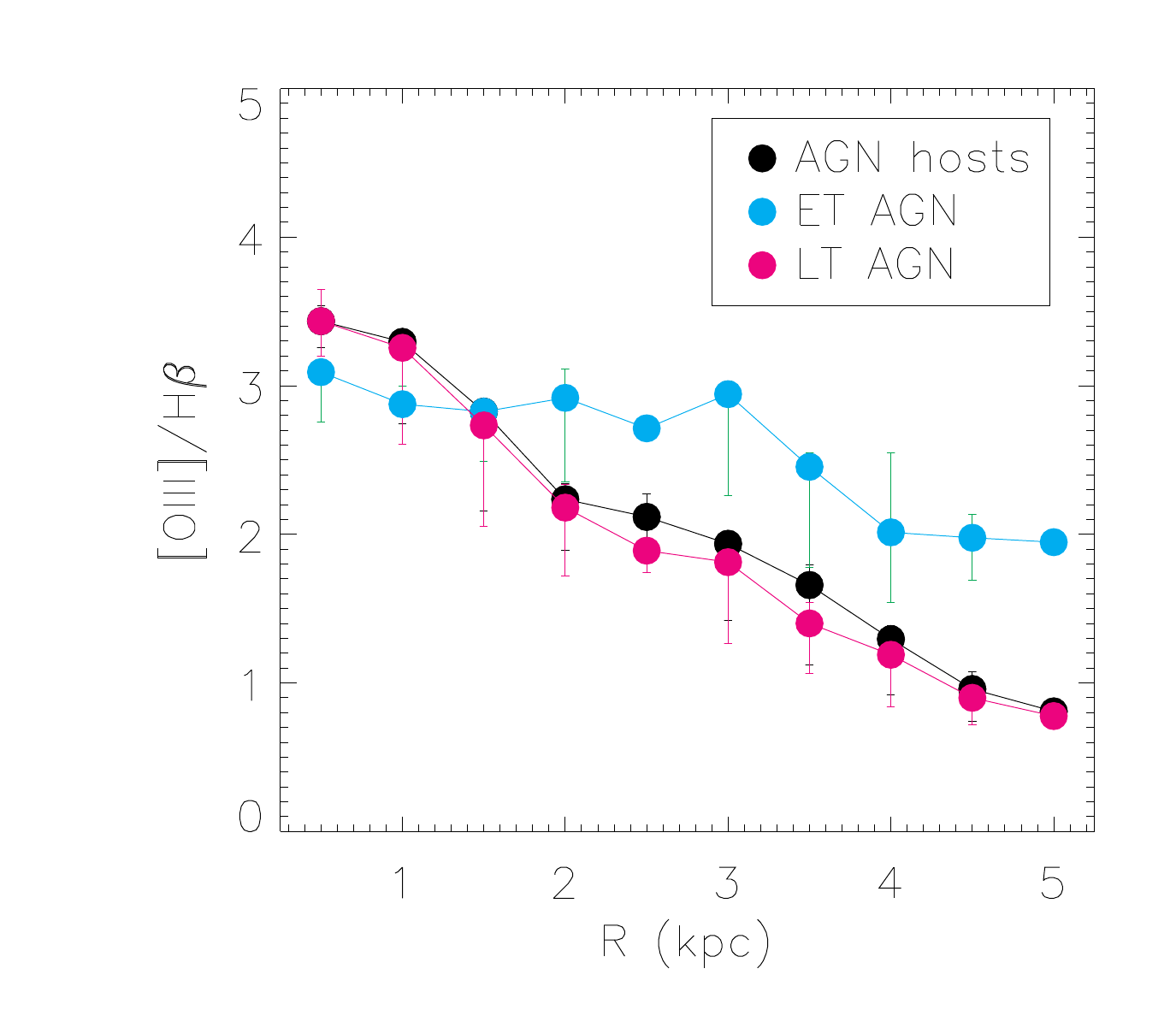}
    \caption{{Comparison of the $[\textsc{Oiii}]\lambda 5007/\textrm{H}\beta$ line ratio spatial profiles. Each point represents the azimuthal median value of the quantities in a bin of 0.5 kpc. The error bars present the standard deviation of the quantities, divided by the square root of the number of spaxels. \textit{Left panel:} AGN hosts (black) and controls (green); \textit{central panel:} strong (dark blue) and weak (red) AGN; \textit{right panel:} early-type (light blue) and late-type (pink) AGN.}}
    \label{fig:OIII_HB}
\end{figure*}

\begin{figure}[h!]
    \centering
    \includegraphics[scale=0.3]{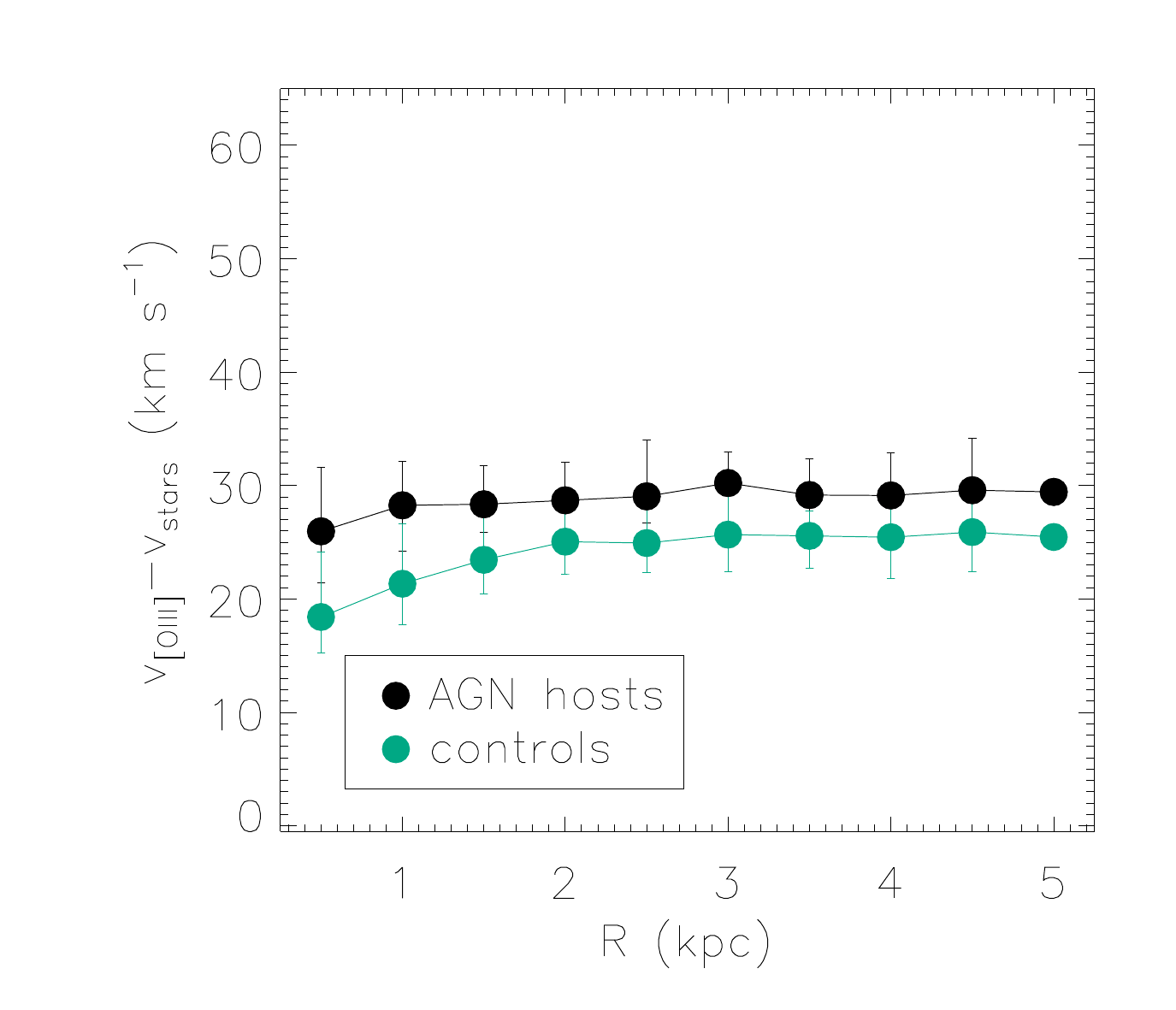}
    \includegraphics[scale=0.3]{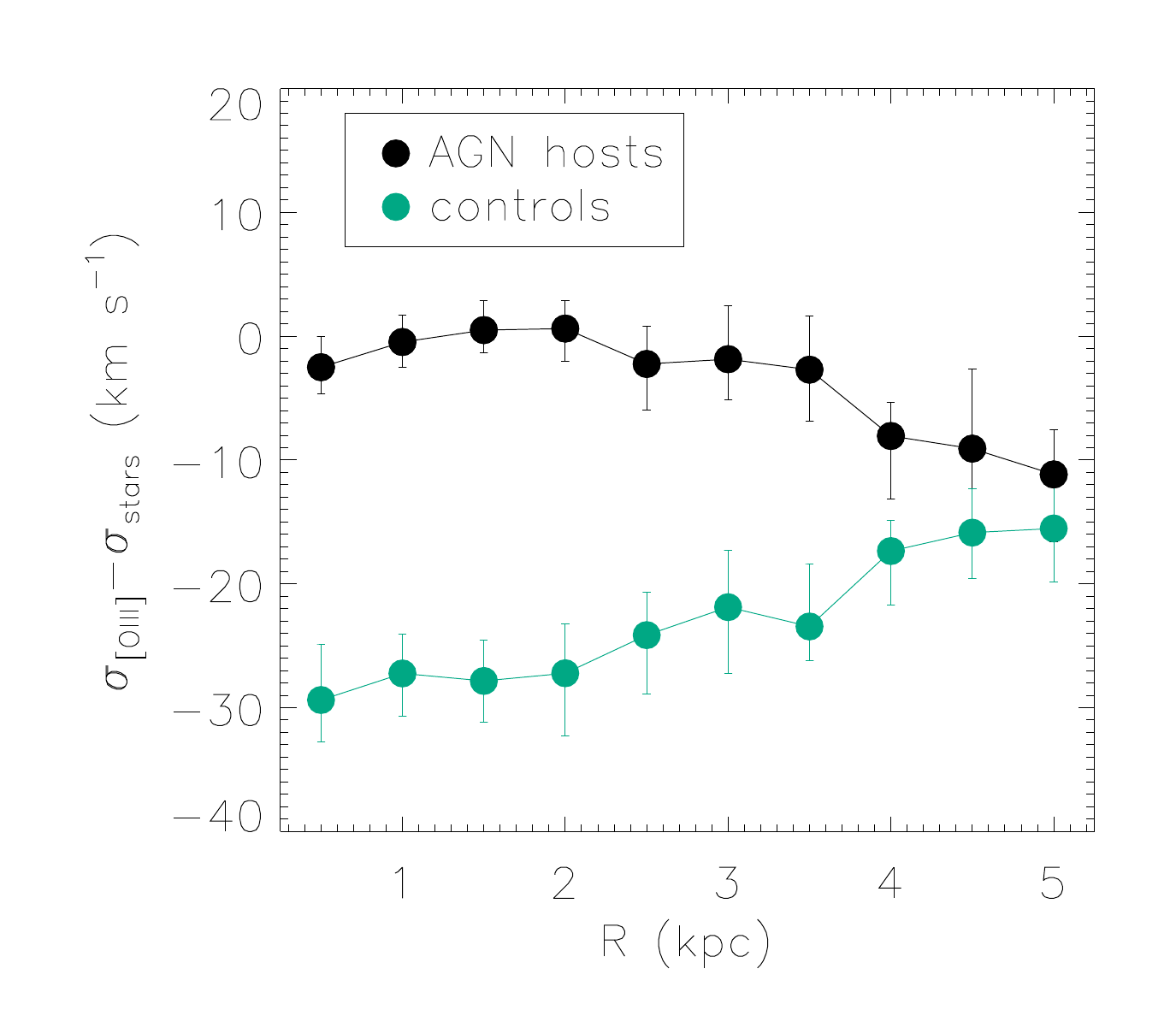}
    \\
    \includegraphics[scale=0.3]{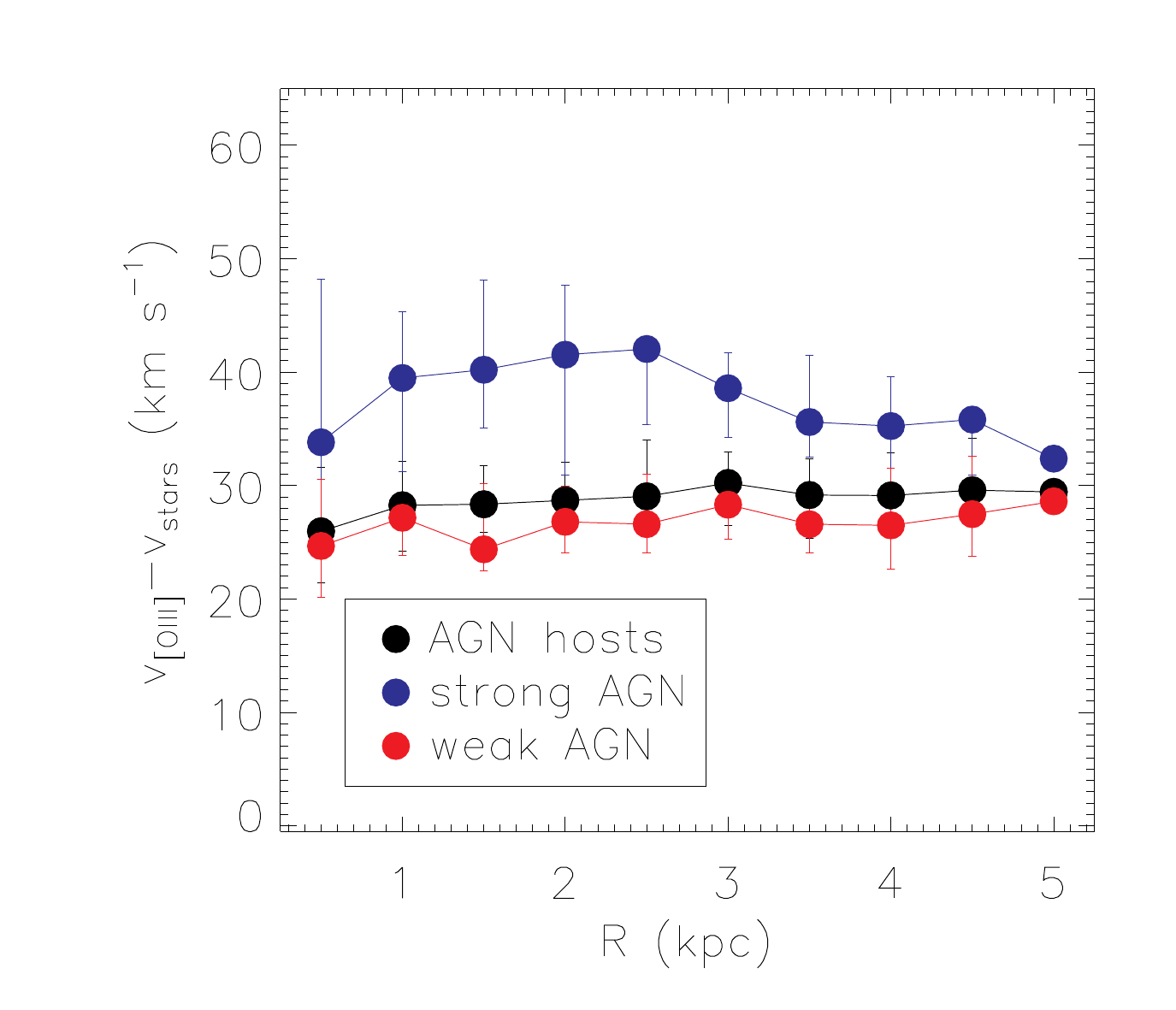}
    \includegraphics[scale=0.3]{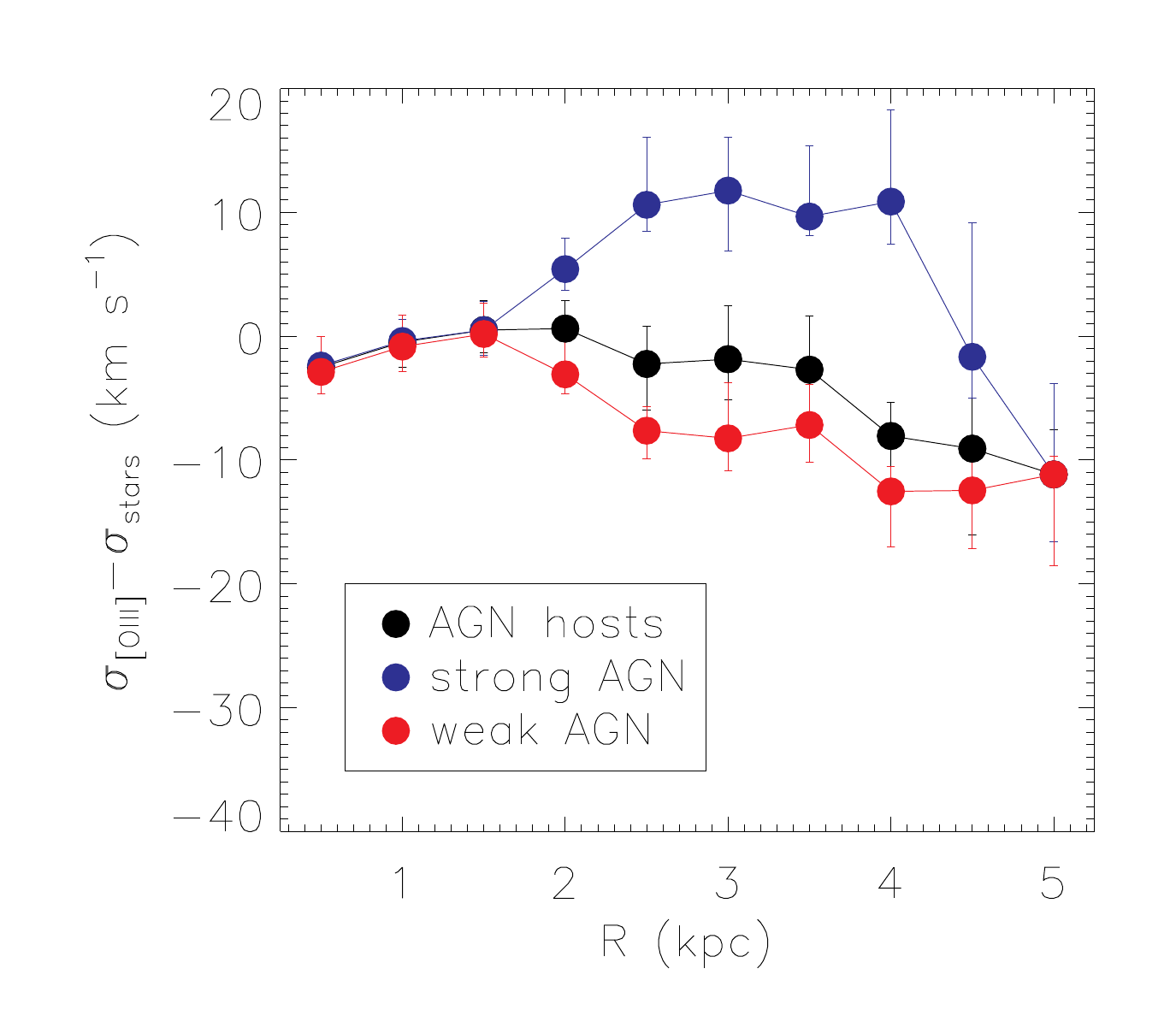}
    \includegraphics[scale=0.3]{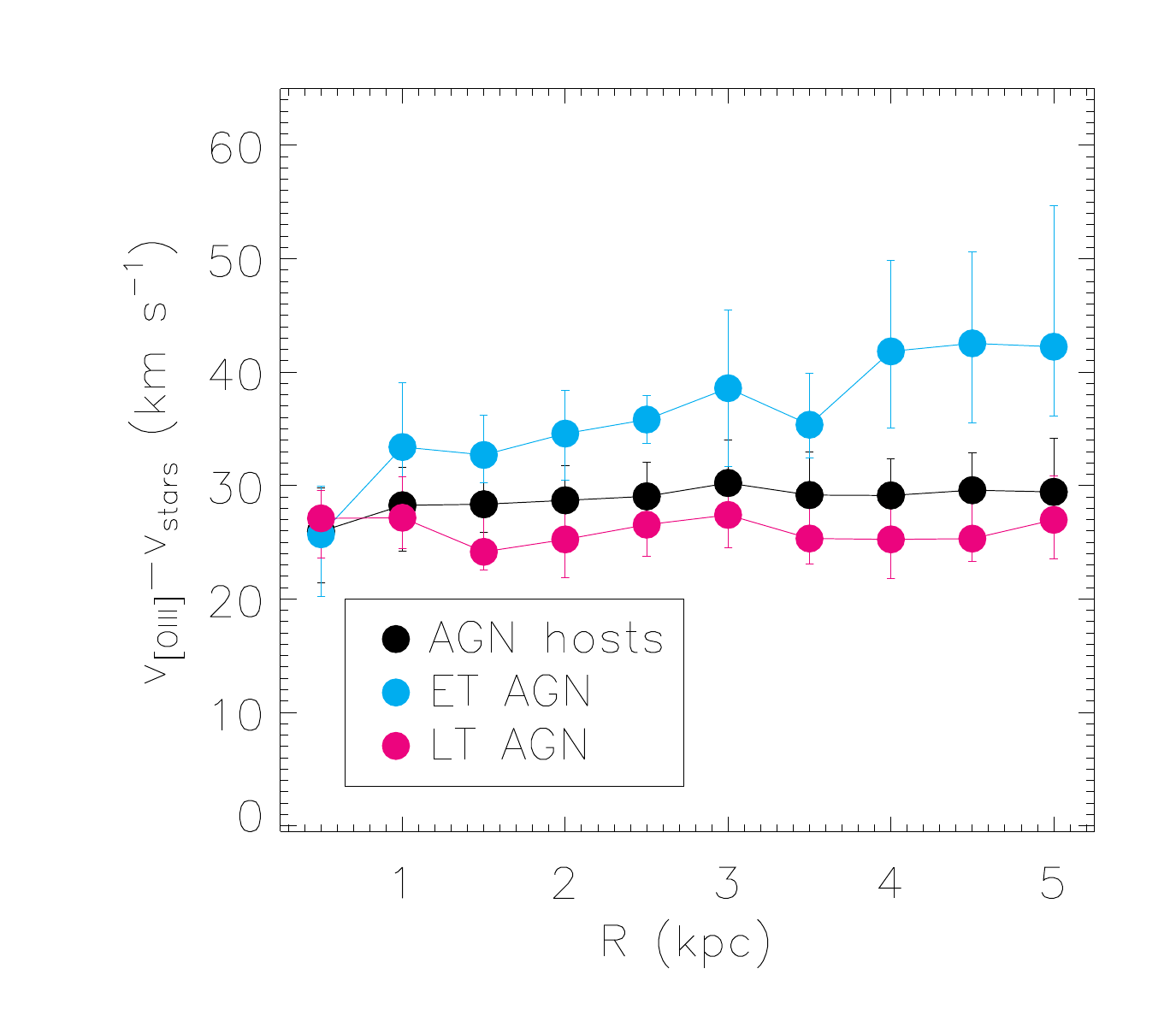}
    \includegraphics[scale=0.3]{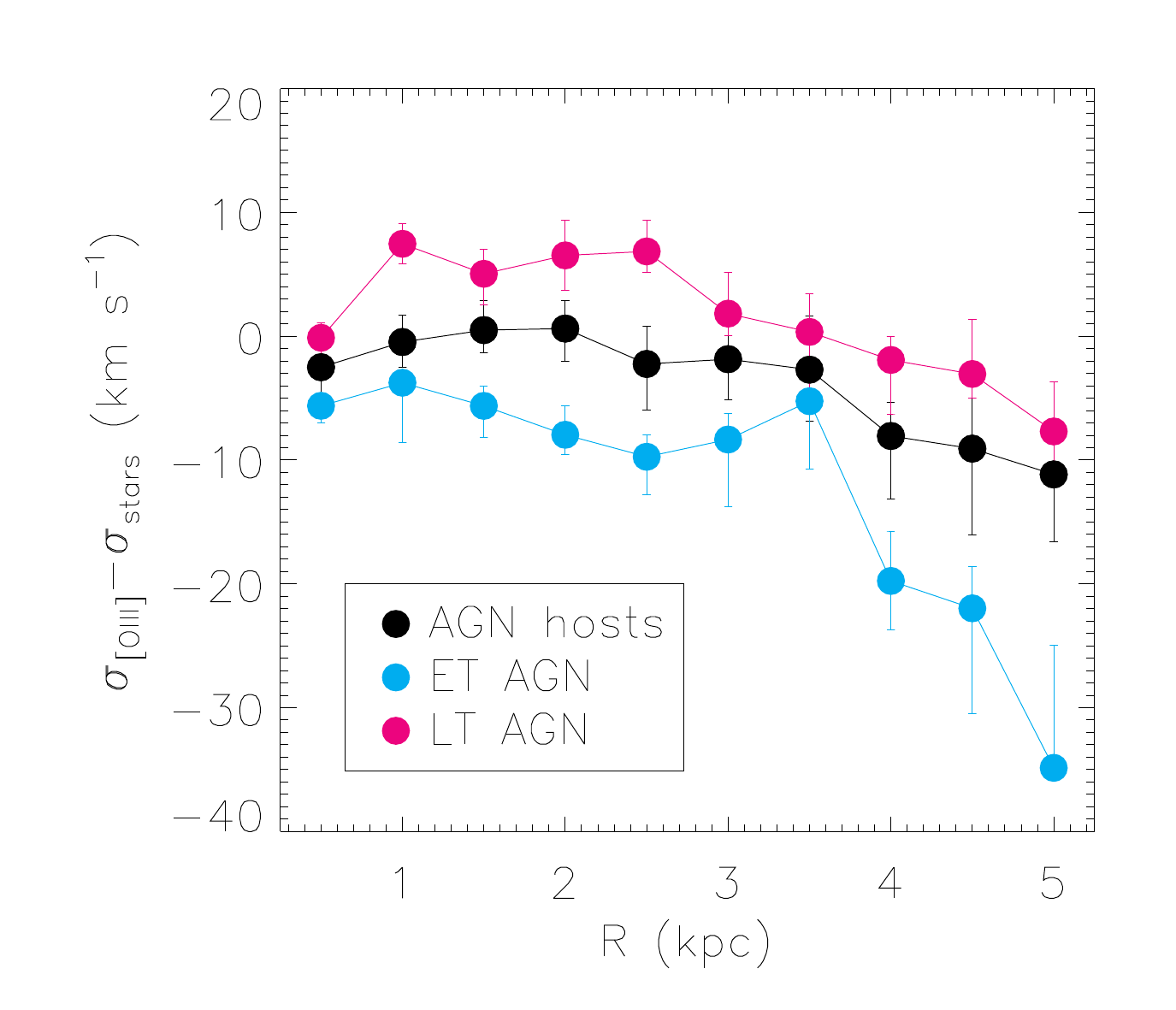}
    \caption{Comparison of median kinematic radial profiles. The symbols and colors are identical to those of Fig. \ref{fig:OIII_HB}. The first row shows a comparison between AGN and controls, from left to right: residual between the $[\textsc{Oiii}]\lambda 5007$ gas and stellar velocities and residual between the corresponding velocity dispersions. Second row: same as in first row but comparing strong and weak AGN hosts. Third row: same as first but comparing early and late-type AGN hosts.}
    \label{fig:samples_profiles_host}
\end{figure}

\begin{figure}
    \centering
    \includegraphics[scale=0.55, trim=1cm 0cm 0cm 1cm, clip]{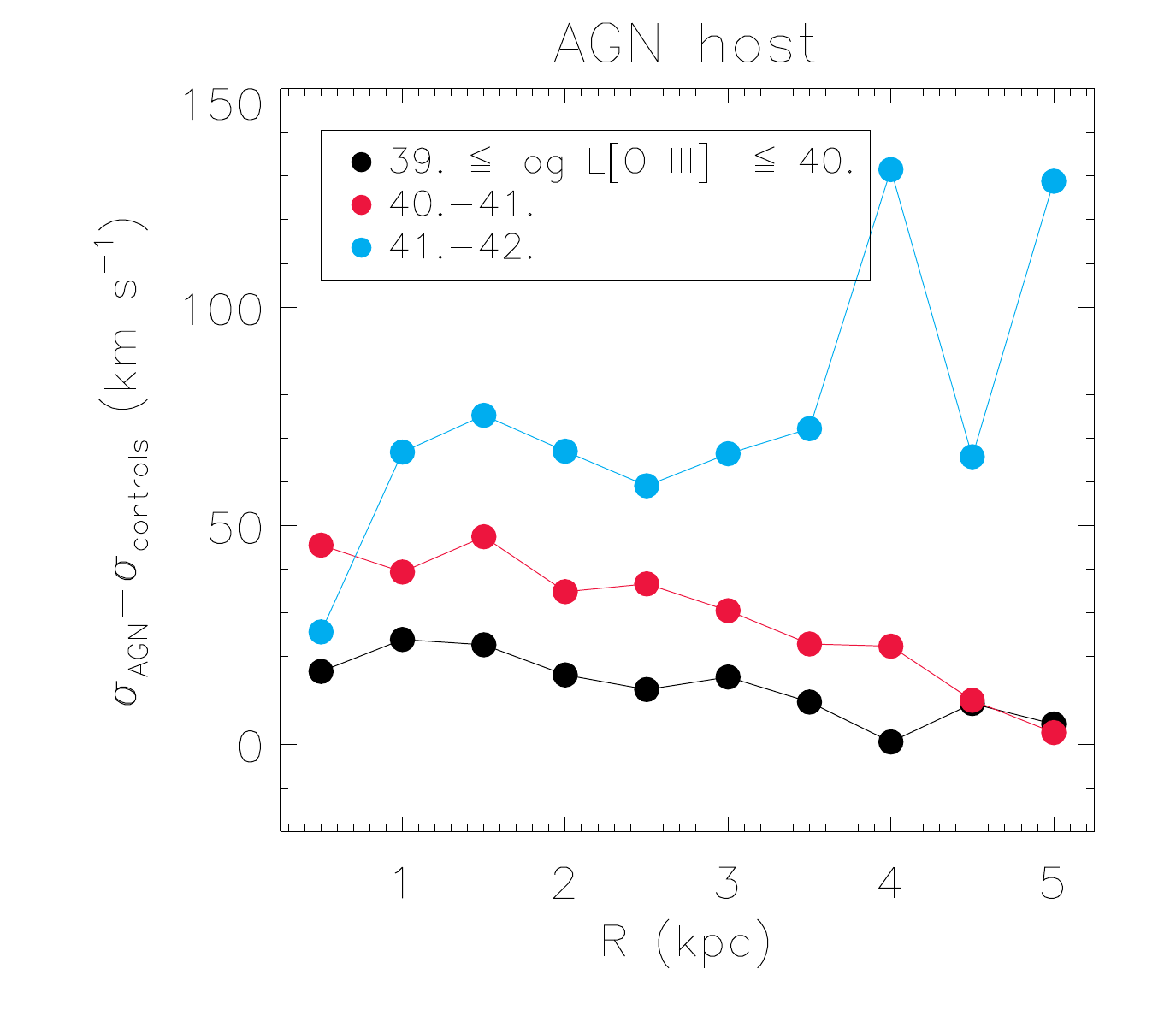}
    \caption{Median radial profile of the $\sigma$ difference between AGN and controls. Each AGN was compared to its respective control galaxies. The colored lines indicate different bins of $[\textsc{O iii}] \lambda 5007$ luminosity, on a logarithmic scale. Over all the analysed extension, the largest differences between AGN and control velocity dispersions were observed for the most luminous AGN hosts.}
    \label{fig:samples_lumin_bpt_host}
\end{figure}

\par  To identify and characterise the kinematically disturbed gas by the AGN, we constructed radial profiles of the gas kinematic properties and emission-line ratios. Due to the MaNGA's spatial resolution and coverage, the radii range that we consider in the present work ranges from $0.5$ to $5.0$ kpc. We preferred not to correct the radial profiles for projection along the azimuthal direction, because this correction is not clear for many galaxies of our sample. Any bias resulting from this is nevertheless compensated by the fact that the control sample galaxies were selected to present the same aspect ratio of the corresponding AGN. For each galaxy, we calculated the azimuthal median of all values whose spaxels are located within a 0.5 kpc bin, for which the uncertainty in [O{\sc iii}]5007 flux is smaller than 30\% and uncertainties in velocity dispersion and centroid velocity are lower than 30  km\,s$^{-1}$. Throughout this article, median profiles for properties of the different sub-samples were constructed by computing the median azimuthal values determined within each radial bin. The residual values ([O{\sc iii}] $-$ stars) were calculated at each spaxel and then the median-valued in each radial bin.  The differences between the kinematic properties of AGN and each control galaxy were computed between their azimuthal median profiles.

\section{Results}\label{results}

\subsection{Excitation} 

{ \par Figure \ref{fig:OIII_HB}   shows median radial profiles of emission-line ratios  for the AGN and control samples. We also show 
bars corresponding to plus and minus the root mean square standard deviation of the quantities, divided by the square root of the number of spaxels. 
These profiles were used to compare the radial behaviour of the $[\textsc{O iii}]\lambda 5007/\textrm{H}\beta$ line ratio of different sub-samples: besides the comparison between AGN and control sample, the panels of Fig. \ref{fig:OIII_HB} also include the AGN profiles separated in strong and weak sub-samples, as well as in sub-samples of early and late-type host galaxies. 
\par Both AGN and control samples' $[\textsc{O iii}]\lambda 5007/\textrm{H}\beta$ ratio profiles decrease with the distance from the nucleus, with a steeper gradient seen for AGN. Although a few control galaxies do not show gas emission, most of them show some emission, but the median  $[\textsc{O iii}] \lambda 5007/$H$\beta$ ratio becomes higher than 1 only in the 1 kpc nuclear bin, while AGN show a median nuclear $[\textsc{O iii}] \lambda 5007/$H$\beta > 3$. Strong AGN display higher $[\textsc{O iii}]/$H$\beta$ ratios for radial distances up to 5\,kpc than weak AGN, indicating that the AGN has a major role in the gas excitation even at large radii. The comparison between early and late-type galaxies reveals that they show similar $[\textsc{O iii}]/$H$\beta$  values at the nucleus, but at distances of 1.5 to 5.0 kpc there are clear differences. 
While the early-type AGN have $[\textsc{O iii}] \lambda 5007/$H$\beta\approx3$ over all distances, this ratio decreases for late-type galaxies with increasing distance from the nucleus. Besides probable H\,{\sc ii} regions dominating the outer regions of the disc of late-type galaxies, another possible explanation for this difference is that, as the amount of gas in early type galaxies is smaller than in late-type galaxies, the AGN radiation reaches greater distances from the nucleus.}

\subsection{Kinematics}
\par Figure \ref{fig:samples_profiles_host} shows the radial profiles of the kinematic properties of the samples obtained from measurements of the  $[\textsc{Oiii}]\lambda 5007$ and stellar kinematics. 
We analyse residual gas velocities and residual velocities dispersion. The residual velocity is defined as the median of the absolute value of $[\textsc{Oiii}]\lambda 5007$ velocity minus the velocity of the stars ($v_{res}=|v_{[\textsc{Oiii}]}-v_{\textrm{stars}}|$). Likewise, the residual velocity dispersion ($\sigma_{\textrm{res}}$) is defined as $\sigma_{[\textsc{Oiii}]}-\sigma_{\textrm{stars}}$, where $\sigma$ is the velocity dispersion. Comparing AGN hosts and controls, the AGN present higher values of these properties, especially if we consider only the strong AGN. All along the analysed extension, $v_{\textrm{res}}$ for AGN hosts is slightly higher than that for the control galaxies; in strong AGN, $v_{\textrm{res}}$ is about 10 km\,s$^{-1}$ larger than in weak AGN.  
Late-type AGN display similar kinematic properties to that of the control and weak AGN samples. Conversely, early-type AGN present higher  $v_{\textrm{res}}$ than late-type AGN, increasing with the distance from the nucleus. 

\par The $\sigma_{\textrm{res}}$ profiles shown in the central column panels of Fig. \ref{fig:samples_profiles_host} indicate that the velocity dispersion of the control galaxies' gas is always lower than that of the AGN sample. In the nuclear aperture, the controls display $\sigma_{\textrm{res}}\approx -25.0$ km s$^{-1}$ whilst for AGN the median value is $\sigma_{\textrm{res}} \approx 2.0$ km s$^{-1}$. Separating the AGN in strong/weak, the highest $\sigma_{\textrm{res}}$ values are dominated by strong AGN all along the analysed extension. While weak AGN do not present positive $\sigma_{\textrm{res}}$, the strong AGN present it out to 4 kpc from the nucleus. Regarding the early and late-type AGN $\sigma_{\textrm{res}}$ profiles, late-type galaxies show the highest residual velocity dispersion, about $8.0$ km s$^{-1}$ in the nuclear aperture. Early-type galaxies yield negative values of $\sigma_{\textrm{res}}$ in all the analysed bins of radius. 

\par As seen in Fig.~\ref{fig:samples_profiles_host}, the main difference in the kinematics of the AGN and control sample is  in the velocity dispersion. A direct comparison of the [\ion{O}{III}]$\lambda$5007 $\sigma$ values of AGN and their controls is shown in Fig. \ref{fig:samples_lumin_bpt_host}. We obtain the median values of the profiles $\sigma_{\textrm{AGN}}-\sigma_{\textrm{controls}}$ in three different bins of luminosity: $39.0\le \log$ L$[\textsc{O iii}] < 40.0$ (hereafter low), $40.0\le \log$ L$[\textsc{O iii}] < 41.0$ (hereafter intermediate), and $41.0\le \log$ L$[\textsc{O iii}] < 42.0$ (hereafter high). Each AGN was compared to its respective control galaxies. We define $\sigma_{\textrm{controls}}=(\sigma_{\textrm{control1}}+\sigma_{\textrm{control2}})/2$. Low and intermediate luminosity AGN present similar profiles, reaching values of $\sigma_{\textrm{AGN}}-\sigma_{\textrm{controls}}\sim20.0$ km s$^{-1}$ and $\sim40.0$ km s$^{-1}$ in the bin corresponding to the inner kpc, respectively.  The most luminous AGN display the highest values of $\sigma_{\textrm{AGN}}-\sigma_{\textrm{controls}}$ in all the radial bins. In the inner 1 kpc, this difference reaches $\approx 65.0$\,km\,s$^{-1}$, indicating a relation between the $[\textsc{O iii}] \lambda 5007$ luminosity and $\sigma$.

\section{Discussion}\label{disc}
By comparing the gas kinematics of AGN hosts and control galaxies, we find evidence that the AGN affect the gas motions up to 4-5 kpc, most clearly seen as differences in the gas velocities dispersion of AGN and controls, but also observable in the centroid velocities. We attribute these differences to AGN driven winds interacting with the NLR gas; in this section, we determine the physical extent of these kinematically disturbed regions (KDRs), the outflow powers and discuss their implication for AGN feedback processes. The KDR is defined here as the region where the gas kinematics is disturbed by the AGN. 

\subsection{Sizes of the NLR and KDR}

\begin{figure}
\centering
\includegraphics[width=0.44\textwidth]{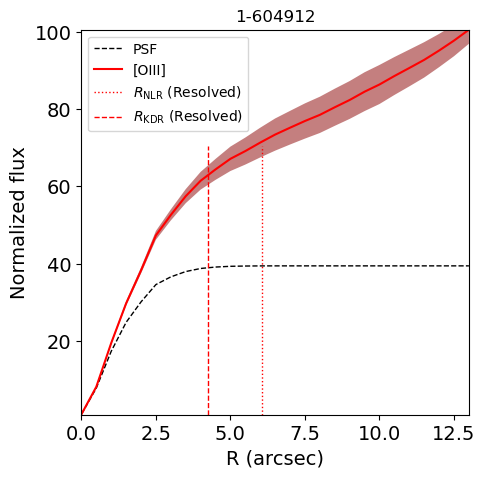}
\includegraphics[width=0.44\textwidth]{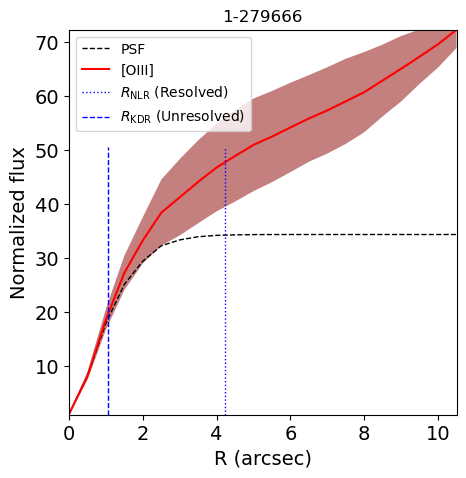}
\caption{Examples of curves-of-growth. The black dashed line corresponds to the PSF and the red line shows the [\ion{O}{iii}]$\lambda$5007 flux and curve-of-growth. The shaded regions delineate the 1$\sigma$ flux uncertainties. The vertical dotted and dashed lines show the observed radii of the NLR and KDR, respectively. The {\it mangaid} of the galaxies are identified in the title of each panel. }
\label{fig:psf}
\end{figure}

\begin{figure}
{\centering
\includegraphics[width=0.44\textwidth]{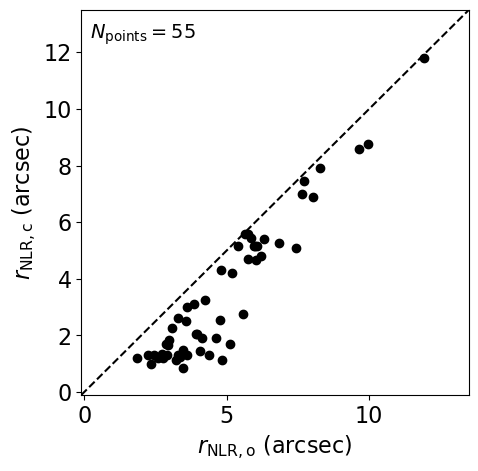}
\includegraphics[width=0.44\textwidth]{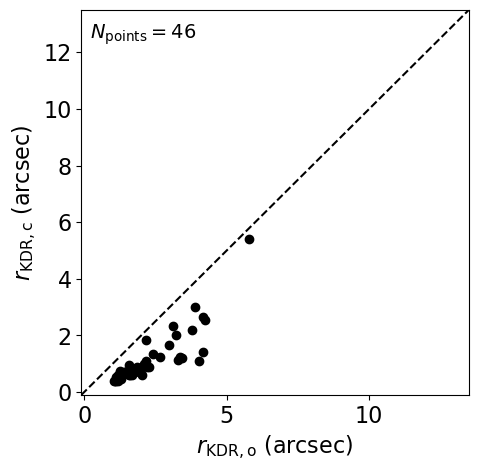}
}
\caption{Comparison of the raw estimates of the
extents of the NLR (left) and KDR (right) with the corrected sizes obtained after  deconvolving the [\ion{O}{iii}] brightness profiles (see text). The dashed lines indicate the locus
of the 1:1 relation.}
\label{fig:deconvolution}
\end{figure}

\subsubsection{Estimates of the NLR and KDR extents}
\begin{figure}
    \centering
    \includegraphics[width=\columnwidth, trim=1cm 0cm 0cm 0cm, clip]{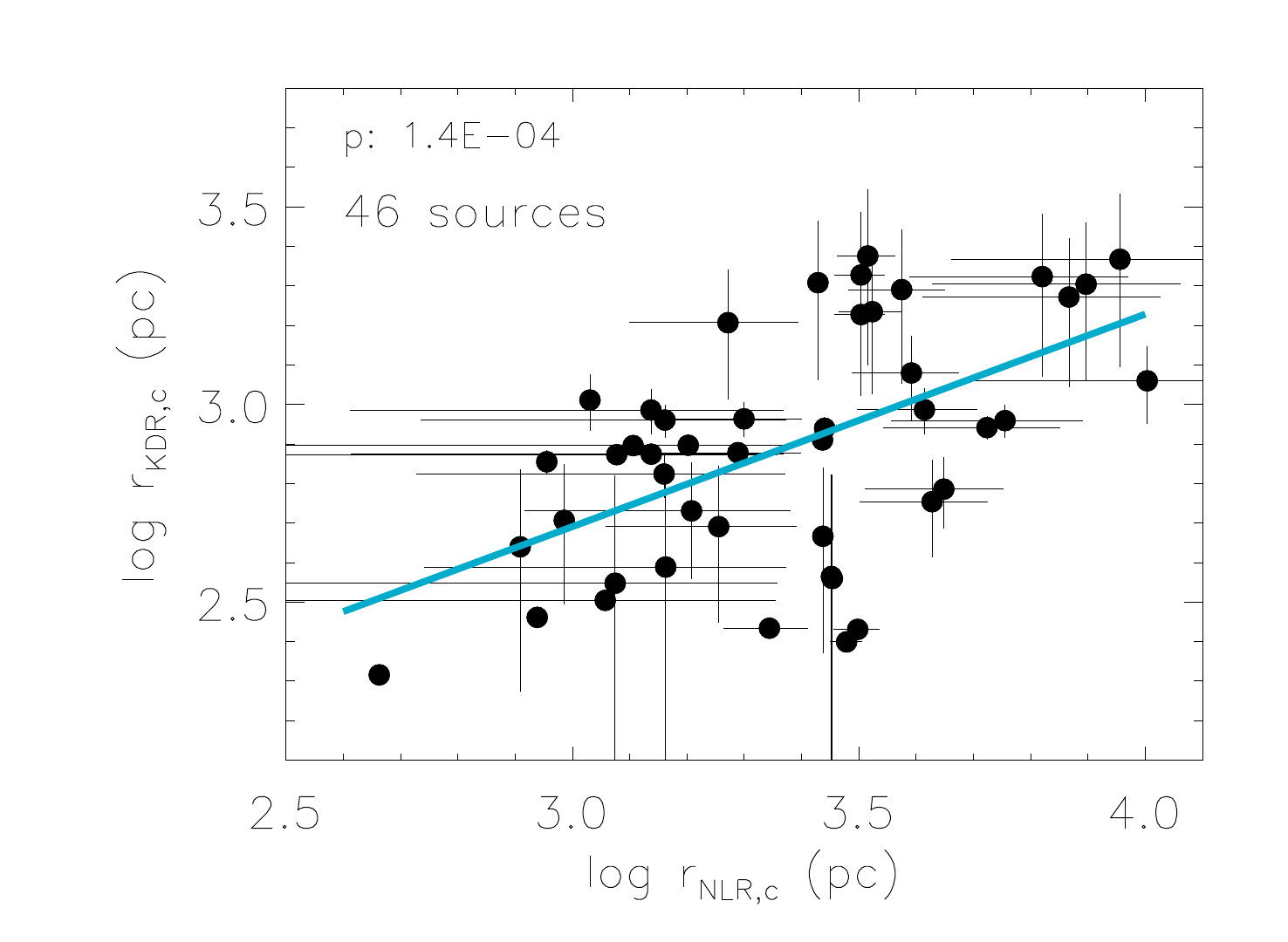}
    \caption{The KDR extents versus the NLR extents on a logarithmic scale. There is a correlation between these extents, with a Spearman correlation coefficient of $CC \sim 0.53$ and a p-value of $10^{-4}$. The best-fit linear regression (blue line) results in an intercept of $\sim1.07 \pm 0.22$ and a slope of $\sim0.53 \pm 0.12$. The uncertainties are represented by gray bars corresponding to differences in the value of the properties of 1$\sigma$.}
    \label{fig:nlr_outflow}
\end{figure}

Measurements of the extent of the NLR in AGN are commonly reported in the literature, mainly based on imaging and Integral Field Spectroscopy (IFS) observations. The extent of the NLR correlates with the AGN luminosity and ranges from a few hundred parsecs in low luminosity AGN  \citep[e.g.][]{fraquelli2003,schmitt2003,nascimento2019,chen2019} to tens of kpc in the most luminous objects \citep[e.g.][]{bennert2002,greene2011,liu2013,liu2014,thaisa2018}. The KDR extents and how they relate to the size of the NLR is less clear. So far, studies on individual objects and small samples find ionised gas outflows on scales from tens of parsecs to kiloparsec \citep{das2005,das2007,thaisa2010,liu2013b,liu2014,barbosa2014,davies2014,cresci2015,rogemar_mrk79,rogemar_n5643,fischer2017,revalski2018,fischer18,diniz2019,munoz-vergara,soto-pinto2019}. Here, we measure both the extent of the KDR and the size of the NLR.

\par We compute the observed extents of the NLR ($r_{\mathrm{ NLR,o}}$) and KDR ($r_{\mathrm{KDR,o}}$) of our AGN sample, by comparing emission-line ratios and gas kinematics of AGN hosts and control galaxies in bins of 0.5 kpc, although the minimal region considered has a size of two spaxels, which corresponds on average to $\sim$ 0.876 kpc in our sample. All the bins take into account only the sources that have measurements in the corresponding region, e. g., some galaxies do not have the first 0.5 kpc resolved and as consequence are not considered in this bin. We define $r_{\mathrm{NLR,o}}$ as the farthest radius from the nucleus within which both $[\textsc{O iii}]/$H$\beta$ and $[\textsc{N ii}]/$H$\alpha$ emission lines ratios fall in the AGN region of the $[\textsc{O iii}]/$H$\beta$ vs. $[\textsc{N ii}]/$H$\alpha$ BPT diagram \citep{bpt}. In addition,  the H$\alpha$ equivalent width is required to be larger than 3.0 $\AA$ to exclude regions ionised by post-AGB stars \citep{stasinska08,cid11,singh13,belfiore16}.

The observed extent of the KDR $r_{\textrm{KDR,o}}$ is  defined as the largest radius from the nucleus where the residual $\sigma_{\textrm{res, AGN}}=\sigma_{\textrm{[OIII], AGN}}-\sigma_{\textrm{stars, AGN}}$ and  becomes equal to $\sigma_{\textrm{res, controls}}=\sigma_{\textrm{[OIII], controls}}-\sigma_{\textrm{stars, controls}}$.  To compare these residuals in both AGN and controls, we fit their radial distribution by a second order polynomial and compute the radius  where they intercept, which is $r_{\textrm{KDR,o}}$.
The extents $r_{\textrm{NLR,o}}$ and $r_{\textrm{KDR,o}}$ could not be determined for the entire AGN sample, since it was not possible to obtain all the necessary measurements (e.g., when one or both line ratios are AGN-like only for the nuclear aperture, or $\sigma_{\textrm{res, AGN}}$ and $\sigma_{\textrm{res,controls}}$ are already similar at the nucleus).

\subsubsection{Are the NLR and KDRs resolved?}
\label{nlrresolved}

The emission-line flux distributions of an unresolved nuclear source can be smeared by the seeing to distances much larger than the FWHM of the point spread function. 
In order to verify whether the NLR and KDR radii $r_{\textrm{NLR,o}}$ and  $r_{\textrm{KDR,o}}$ derived in the previous section are spatially resolved, we follow \citet{kakkad20} and compute the [\ion{O}{iii}]5007 emission-line flux curve-of-growth (CofG) for each galaxy. We then compare the [\ion{O}{iii}]5007 flux and PSF CofGs, with the latter obtained from the reconstructed g-band point source profiles included in the MaNGA data-cubes. The CofG are not used to estimate the  NLR or KDR radii, but just to check if they are spatially resolved by MaNGA. The [\ion{O}{iii}]5007 emission line is chosen for this purpose because it is  the best tracer of the gas ionized by the AGN and of the gas in outflow, being one of the strongest optical lines.   
First, we normalise the  [\ion{O}{iii}] flux by  its the value in the nuclear spaxel -- defined as that corresponding to the peak of the continuum emission -- and then compute the integrated fluxes increasing the radius in steps of 0\farcs5. The same procedure is adopted to compute the CofG for the PSF.

In Fig.~\ref{fig:psf} we show examples of the CofGs. We consider that the KDR is spatially resolved if the [\ion{O}{iii}]  flux in the CofG at $r_{\rm KDR,o}$ is higher than that of the PSF considering the 1$\sigma$ error (i.e., the shaded red region must be above the PSF CofG profile). Similarly, the NLR is considered resolved if at $r_{\rm NLR,o}$ the [\ion{O}{iii}] flux of the CofG is larger than that of the PSF CofG.   The left panel of Fig.~\ref{fig:psf} shows an example in which both $r_{\rm KDR,o}$ and $r_{\rm NLR,o}$ are spatially resolved,
while the right panel shows an example in which the KDR is unresolved and the NLR is resolved. We find that $r_{\rm NLR,o}$ and $r_{\rm KDR,o}$  are spatially resolved in 55  and 46 AGN hosts of our sample, respectively. Our further analysis will be focused on these objects.

\subsubsection{Correction for the beam smearing effect}

Having established that the NLR and KDR of most AGN in our sample are spatially resolved, we now correct their measured extents for beam smearing. In particular, our estimates of the extent of the KDRs are based on kinematic parameters, which are hard to correct using purely photometric techniques like surface brightness deconvolution. In what follows, we simply assume that beam smearing affects equally  the derived linear  extents obtained  from the kinematic and photometric analyses, $r_{\textrm{KDR,o}}$ and $r_{\textrm{NLR,o}}$, respectively. This assumption allows us to perform a rough correction to the estimated radial extents.

Our methodology for estimating the corrected extents of both the NLR ($r_{\textrm{NLR,c}}$) and KDR ($r_{\textrm{KDR,c}}$) is as follows. We start by deconvolving the [O\,{\sc iii}]5007 brightness profiles of each galaxy, which were derived by calculating the integrated fluxes in concentric rings around the galaxy centre at fixed steps 0\farcs5 wide. The same calculation was performed for the PSF associated to each galaxy (see Sec. \ref{nlrresolved}). We used these profiles to derive 2D circularly symmetric flux distributions for [\ion{O}{iii}] and the PSF with a cubic spline fit, producing ``pseudo-images'' whose linear scale was lower than the original MaNGA data-cubes by a factor 10. We then performed a 2D Richardson-Lucy deconvolution on the [\ion{O}{iii}] brightness distributions using the Python package {\sc{scikit-image}}, with the 2D symmetric PSF as input. Finally, we calculated in the observed profile the fraction of the total [\ion{O}{iii}] flux contained within the raw, uncorrected radial extent of a given structure (KDR or NLR), and impose that the corrected extent comprises the same flux fraction in the deconvolved profile.

Fig.~\ref{fig:deconvolution} shows the comparison between the raw estimates
of the extent of the KDR/NLR and those corrected as explained above. The beam
smearing, as expected, produces an  overestimation of the true extent of both structures.
In some cases, particularly for  the most extended ($\gtrsim 5^{\prime\prime}$) sources, the corrected
sizes are very similar to the raw ones, which is also expected. The mean ratio between the corrected and observed radii are 0.48$\pm$0.14 and 0.65$\pm$0.22  for $r_{\textrm{KDR}}$ and $r_{\textrm{NLR}}$, respectively.

\subsubsection{Comparison between the NLR and KDR extents}

Fig. \ref{fig:nlr_outflow} displays the $r_{\textrm{KDR,c}}$  against the $r_{\textrm{NLR,c}}$. The average $<r_{\textrm{NLR,c}}>$ is $3.00\pm0.33$ kpc and individual values range between $0.4$ and $10.1$ kpc, which are in good agreement with previous values obtained for MaNGA AGN based on spatially resolved BPT diagrams \citep{chen2019}. The KDRs present extents of: $<r_{\textrm{KDR,c}}>=0.96\pm 0.09$\,kpc, varying between 0.2 and 2.3 kpc. \citet{wylezalek2020} found that the average width of the  [OIII]$\lambda5007$ line in weak AGN reaches the same level of non-AGN at distances of $\sim$8\,kpc from the nucleus, while for strong AGN similar values are seen at distances of up to 15\,kpc, leading to a `qualitative' measurement of the extent of the KDR that are on average larger than ours. The p-value of the correlation between $r_{\textrm{KDR,c}}$ and $r_{\textrm{NLR,c}}$ of our sources is $10^{-4}$, suggesting that they are correlated.
We fit the data by a linear equation and find
\begin{equation}
\log r_{\rm KDR,c} = (0.53\pm 0.12)\,\log r_{\rm NLR,c} +(1.07\pm 0.22)\,, 
\end{equation}
\noindent in units of kpc. $r_{\textrm{KDR,c}}$ and $r_{\textrm{NLR,c}}$ present a positive correlation and, on average, $r_{\textrm{KDR,c}}$ is smaller than  $r_{\textrm{NLR,c}}$, with mean value of $\langle r_{\textrm{KDR,c}}/r_{\textrm{NLR,c}}\rangle = 0.32\pm0.27$. Using HST observations of 12 nearby type 2 quasars ($z<0.12$, $L_{\rm bol}\sim10^{46}$ erg s$^{-1}$), \citet{fischer18} reported that the extent of the outflows are on average only 20\,\% of the extent of the  [O\,{\sc iii}] emission (NLR or ENLR) in a sample of nearby type 2 quasars. Other studies on quasars at larger distances and luminosities ($z=0.5-0.6$, $L_{\rm bol}\sim10^{47}$ erg s$^{-1}$) using integral field spectroscopy and HST images find  similar extents for the outflow and ionised gas emission \citep{liu2013b,liu2014,wylezalek2016}. As a cautionary note, we point out that the methods used to estimate the size of the KDR and outflows vary widely among all these studies and the comparison between different measurements is not straightforward.

\subsection{Mass outflow rates} 
\begin{figure*}
    \centering
    \includegraphics[scale=0.5, trim= 0.9cm 0cm 0cm 0cm, clip]{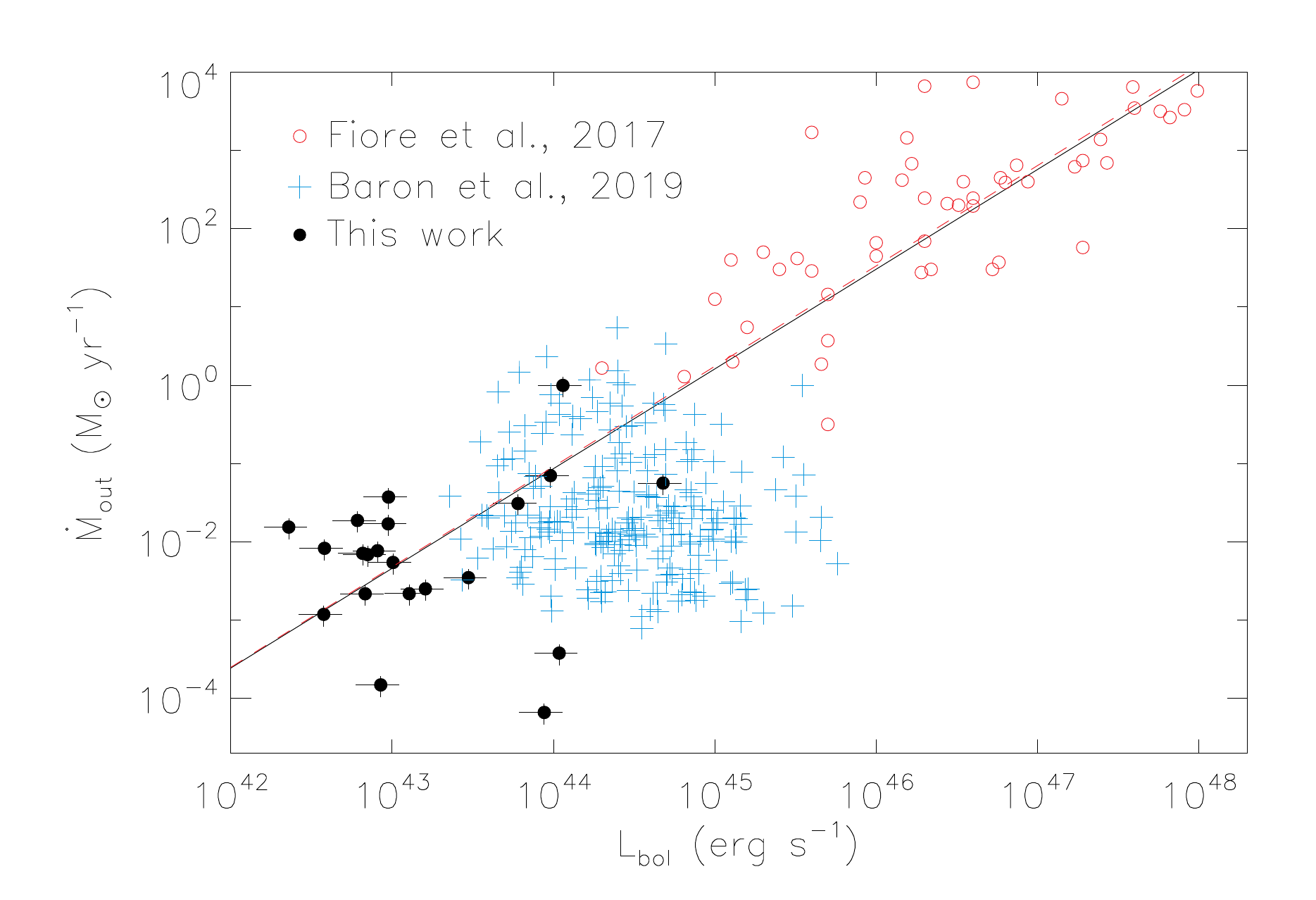}
    \includegraphics[scale=0.5, trim= 0.9cm 0cm 0cm 0cm, clip]{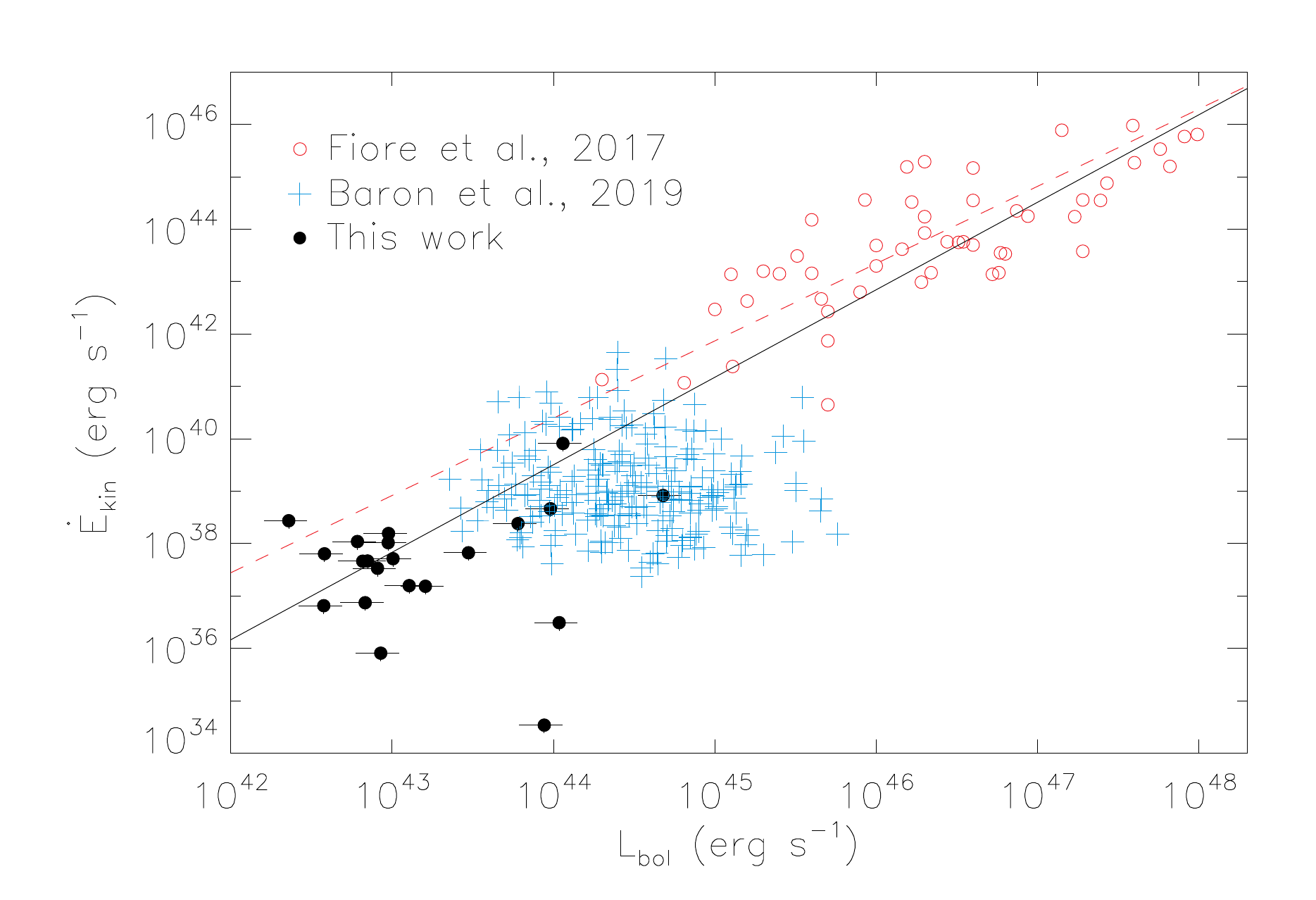}
    \caption{$\dot{M}_{out}$ (left panel) and $\dot{E}_{kin}$ (right panel) versus the AGN bolometric luminosity. Red circles indicate objects analysed in \citet{Fiore}, while blue crosses show the objects from \citet{baron2019b}. Our sample is represented by black dots and characterise AGN with the lowest luminosity in these relations. We were able to estimate these properties for 20 sources. Black lines show the linear fits considering the three samples. The red dashed lines indicate the linear fit presented in \citet{Fiore}.} 
    \label{fig:estimativas}
\end{figure*}

\par In order to examine the outflow impact on their host galaxies, we first must estimate the mass outflow rate. Following previous works (e.g., \cite{rogemar_mrk79}, \cite{riffel2}, \cite{thaisa2010}), we define the mass outflow rate $\dot{M}_{\rm out}$ as 
\begin{equation}\label{eqmout}
\dot{M}_{\rm out}=1.4m_pN_ev_{\rm out}fA ,
\end{equation}
in which $m_p$ is the proton mass, $N_e$ is the electron density, $v_{\rm out}$ is the outflow velocity in the $[\textsc{O iii}]\lambda 5007$ emission line, $f$ is the filling factor, $A$ is the area of the outflow cross section and 1.4 factor is due to the estimated contribution  from He to the gas mass. The filling factor can be estimated by
\begin{equation}
f\approx \frac{L_{\rm H\alpha}}{4\pi j_{\rm H\alpha}V}, 
\end{equation}
where $L_{\rm H\alpha}$ is the H$\alpha$ luminosity emitted by a volume $V$ and $4\pi\,j_{\rm H\alpha}/N_e^2=3.534\times10^{-25}$ erg\, cm$^{-3}$\,s$^{-1}$ \citep{osterbrock06}. Here, we adopt the assumption that all the gas is in outflow, as our data does not allow the separation between gas in the outflow from gas that is not. This assumption may introduce uncertainties that are nevertheless not larger than those introduced by our ignorance of other variables in these calculations, such as the outflow geometry and orientation. Replacing $f$ in Eq.\,\ref{eqmout} and assuming a spherical geometry for the outflow, we obtain
\begin{equation}\label{eq_out}
\dot{M}_{\rm out}=\frac{4.2 m_p L_{\rm H\alpha} v_{\rm out}}{N_e j_{\rm H\alpha} r},
\end{equation}
where $r$ is the radius of the sphere. 

As pointed out in the previous section, the radius of the outflow is considered to be smaller than $r_{\textrm{KDR,c}}$, as previous resolved studies have shown that the mass outflow rates (and powers) \citep[e.g.][]{thaisa2018,fischer2017} usually peak within the inner kpc of the host galaxies and then decrease for larger distances. This ``peak radius" is the one to be used in calculations $\dot{M}_{\rm out}$ and outflow power, although the effect of the outflows (disturbed kinematics) may extend much beyond this radius. Nevertheless, such radii are not  resolved in our MaNGA data. We have thus adopted the outflow radius as the smallest value consistent with our observations, adopted to be the distance corresponding to the radius of the seeing disk of the observation of each source. This value at the galaxies approximately correspond to typical values obtained in resolved studies of $\approx$\,1\,kpc.

The electron density values  were calculated for each galaxy from the $[\textsc{S II}]\lambda 6717/[\textsc{S II}]\lambda 6731$ ratio and assuming a temperature of 10\,000\,K. Recent works have been arguing that the density of the gas in outflow obtained using $[\textsc{S II}]$ is an underestimation for outflows occurring on hundred of pc scales, as the $[\textsc{S II}]$ emission originates from a partially ionised region farther out from the nucleus, that has a lower gas density \citep{baron2019a,baron2019b,davies2020}. But, as the radii of the outflows are of the order of kpc for our sample, the $[\textsc{S II}]$ emission can be used to obtain the gas density at these distances. 

The $[\textsc{S II}]$ doublet is not sensitive to $50 > N_e\gtrsim1\times10^4$\,cm$^{-3}$, where the dependence of the $[\textsc{S II}]$ line ratios with $N_e$ becomes asymptotically flat. Thus, we estimate the outflow properties only in objects with measured densities $50<N_e<1\times10^4$\,cm$^{-3}$. Among the AGN hosts with spatially resolved outflows, we estimate $50<N_e<1\times10^4$\,cm$^{-3}$ for 43 sources. The mass outflow rate and kinetic power of the outflows are estimated only for these objects.
In any case, mass outflow rates $\dot{M_{out}}$ and kinetic power of the outflows $\dot{E_{kin}}$ estimated in this work should be treated as upper limits, the electron density being the major source of uncertainties \citep[e.g.][]{harrison18,davies2020}.

The differences observed in the gas kinematic profiles of AGN hosts and their controls can be attributed to the nuclear activity. We calculate the outflow velocity as
\begin{equation}
v_{\rm out}=|v_{[\textsc{Oiii}],AGN}-v_{stars,AGN}|. 
\end{equation}
We calculated $v_{\rm{out}}$ for each galaxy within an aperture corresponding to the MaNGA angular resolution (2\farcs5 FWHM)  -- adopted as the outflow radius -- and then compared each AGN with its respective control galaxy by subtracting their velocities.

The median mass outflow rate of our AGN sample is $\dot{M}_{\rm out}= 7.1 \times 10^{-3}$ $M_{\odot}$ yr$^{-1}$. This value is highly uncertain, mainly due to the assumptions regarding the geometry of the outflow and its density. However, by adopting the same assumptions for the whole sample, we can compare the derived mass-outflow rates of all AGN host galaxies. This is done in Fig. \ref{fig:estimativas} that presents $\dot{M}_{\rm out}$ vs. the AGN bolometric luminosity ($L_{\rm bol}$). The $L_{\rm bol}$ was obtained using the following relation from \citet{heckman_2004}:
\begin{equation}
    L_{\rm bol}\sim 3\,500 L[\textsc{O iii}],
\end{equation}
\noindent where $ L[\textsc{O iii}]$ is the  [\textsc{O iii}]$\lambda$5007 luminosity measured within a circular aperture of 2\farcs5 diameter.

\par For our sample alone, we do not fin a strong correlation between $\dot{M}_{\rm out}$ and $ L[\textsc{O iii}]$, but the AGN in our sample span a small luminosity range. For comparison, we include in Fig. \ref{fig:estimativas} the measurements of ionised gas outflows from \citet{Fiore} for high luminosity AGN ($L_{\rm bol} \gtrsim 10^{45} {\rm erg s^{-1}}$) and from \citet{baron2019b} for intermediate luminosity AGN ($L_{\rm bol}\approx10^{44}-10^{46} {\rm erg s^{-1}}$). Our data fill the low luminosity end of the relation ($L_{\rm bol}\approx10^{42}-10^{45} {\rm erg s^{-1}}$). The galaxies of our sample are equally spread above and below the relation obtained by \citet{Fiore} for higher luminosity AGN (red dashed line). The galaxies from the sample of \citet{baron2019b} fall mostly below the relation due to their adopted smaller distances and higher densities and thus lower mass outflow rates. 

In Fig.~\ref{fig:estimativas}, we have separated the values of $\dot{M}_{\textrm{out}}$ obtained in the present work, combined with those of \citet{Fiore} and \citet{baron2019b}, in bins of $\log L_{\rm bol}=0.5$ and fit the average values of $\dot{M}_{\rm out}$ in these bins as a function of the AGN luminosity $L_{bol}$ with a linear relation:

\begin{equation}
    \log \dot{M}_{\rm out}=-57.10\pm 0.76 + (1.27\pm 0.11)\log L_{\rm bol},
\end{equation}
 which is shown as a black line in the left panel of Fig. \ref{fig:estimativas}. However, this result must be taken with caution, as the mass-outflow rates are determined with different methods in the three studies, and the scatter of the relation in the low-luminosity end is large \citep{shimizu19}. 

\subsection{Kinetic power of the outflows} 

We now obtain the kinetic power of the outflows $\dot{E}_{\rm kin}$, which is defined as
\begin{equation}
\dot{E}_{\rm kin}\approx\frac{\dot{M}_{\rm out}}{2}(v_{\rm out}^2+\sigma_{\rm out}^2),
\end{equation}
in which $v_{\rm out}$ is the outflow velocity defined in Eq. 5.  We use the [O\,{\sc iii}] velocity dispersion as a proxy of $\sigma_{\rm out}$, which may be a lower limit, as usually the outflows appear as broad wings in the line profiles. We obtain a median value of $\dot{E}_{kin}=5.14\times 10^{37}$ erg s$^{-1}$ for the MaNGA AGN sample. 
The right panel of Fig.~ \ref{fig:estimativas} shows $\dot{E}_{\rm kin}$ vs. $L_{\rm bol}$. As for $\dot{M}_{\rm out}$, we include the compilations from \citet{Fiore} and \citet{baron2019b}. 
Some of our derived values of $\dot{E_{\rm kin}}$ are on and above the extrapolation of the relation of \citet{Fiore}, but most are below,  similarly to the points from \citet{baron2019b}. We obtain the following relation from the fit of the three samples with a linear regression:
\begin{equation}
 \log \dot{E}_{\rm kin}=-31.44\pm 0.51 + (1.59\pm 0.07)\log L_{\rm bol}.
\end{equation}
which is shown as a black line in the right panel of Fig. \ref{fig:estimativas}. We point out that our values of $\dot{E}_{\rm kin}$ may be lower limits, due to our assumed velocity dispersion of the outflow, used in the determination of $\dot{E}_{\textrm{kin}}$.

\par Numerical simulations \citep{hopkins_2010} indicate that AGN feedback is only important if their kinetic efficiency is $\dot{E}_{\rm kin}/L_{\rm bol} > 5 \times 10^{-3}$, and the AGN feedback becomes more effective when $\dot{E}_{\rm kin}$ is at least $\sim$5\,\% of the bolometric luminosity \citep{harrison18}. In our sample, this occurs only for one source (MaNGA ID 1-258599). Therefore, the majority of outflows detected in our AGN sample  are not powerful enough to affect their host galaxies on a large scale. The gas is outflowing from the nucleus and being re-distributed within the galaxies remaining available for further star formation and AGN feeding, comprising the so-called  ``maintenance mode" AGN feedback. However, the kinetic power of the ionised outflows corresponds only to their mechanical effect in a single gas phase and do not correspond to the initial AGN input energy  included in theoretical studies, and thus a comparison between the observed kinetic efficiencies in a specific wind phase and theoretical predictions in not straightforward \citep[e.g.][]{harrison17}.
\par We have also measured the escape velocity (V$_{\textrm{esc}}$) for each galaxy and found that the outflow velocity is lower than the V$_{\textrm{esc}}$ in every source, which reinforces the conclusion that the AGN impact is low, the feedback being only in ``maintenance mode''.

\section{Conclusions}\label{conc}
\par We have compared the ionised gas kinematics of a sample of 170 active galaxies to those of a control sample of non-active galaxies from the MaNGA survey. Our goal was to look for properties related to the nuclear activity. Using these properties, we estimated the extent of the NLR and of the Kinetically Disturbed Region -- KDR (attributed to the interaction of AGN outflows with ambient gas in the host galaxy), showing average kinematic profiles as a function of the radius (in units of effective radius). We also estimated the average values of the mass outflow rate within the inner kpc and the kinetic power of outflows for the AGN sample. Our main conclusions are:
\begin{itemize}

\item We find spatially resolved NLR's and KDR's in 55 and 46 AGN host galaxies, respectively, among the 170 AGN of our sample;
    
\item The main difference in the mean gas excitation of AGN relative to controls was observed in the line ratio $[\textsc{O iii}]\lambda 5007/$H$\beta$, which was used to obtain the extent of the NLR;
    
    \item The main difference in the kinematic radial profiles of AGN relative to controls was observed in the residual velocity dispersion $\sigma_{\textrm{res}}=\sigma_{\textsc{O iii}}-\sigma_{\textrm{stars}}$; 
    
    \item Over the whole analysed radial extent, the most luminous AGN ($41.0 \le \log L[\textsc{O iii}] \le 42.0$) present  the highest residual velocities $v_{res}=|v_{[\textsc{O iii}]}-v_{\textrm{stars}}|$ and residual velocity dispersions $\sigma_{res}$ relative to their respective control galaxies;

     \item The NLR extent -- adopted as the region where the AGN is the dominant source of ionisation -- ranges from 0.4 kpc to 10.1 kpc;

    \item The KDR extent ranges from 0.2 to 2.3 kpc, defined as the largest distance from the nucleus where the residual between $\sigma_{\rm [OIII]}$ and $\sigma_{\rm stars}$ for the AGN and controls become similar. The KDR size is on average 32 percent of the mean NLR extent;

    \item There is a correlation between the extent of the KDR and the extent of the NLR: the best fit is $r_{\rm KDR,c}=(0.53\pm0.12)r_{\rm NLR,c}+(1.07 \pm 0.22)$;
    
   \item Assuming that the kinematic disturbance observed along the NLR is due to outflows from the AGN, we have estimated the values of the mass outflow rate and power adopting as radius of the outflow the angular resolution of the data;
    
    \item The mass outflow rate ranges between $\sim 10^{-5}$ and $\sim 1\,M_{\odot}$\,yr$^{-1}$ and shows a weak correlation with the AGN luminosity;
    
    \item The power of the outflows ranges between $\sim 10^{34}$ and $\sim 10^{40}$ erg\,s$^{-1}$, and presents a weak correlation with the AGN luminosity -- although being probably underestimated -- with most values below those expected from the extrapolation of the \citet{Fiore} relation for more luminous AGN;
    
    \item The ratio between the kinetic power of the outflow and the AGN bolometric luminosity $\dot{E}_{\textrm{kin}}/L_{\textrm{bol}}$ ranges between $\sim 10^{-7}$ and $10^{-4}$. Only for the AGN with MaNGA ID 1-258599, which is one of the most luminous sources in our sample, it is higher than the 0.5\% threshold  AGN feedback  postulated by models to significantly affect the evolution of the host galaxy.

\end{itemize}
Although we have found that  the kinematic disturbance on the NLR by the AGN extends up to several kpc in some galaxies, its effect is not powerful enough to significantly affect their host galaxies.  Nevertheless, other gas phases -- neutral and molecular -- as well as other forms of feedback besides the kinetic one seen in the ionised gas, should also be considered as possible sources of AGN feedback for a fairer comparison with galaxy evolution models.

\
\begin{acknowledgements}
The authors thank the anonymous referee for her/his valu-able suggestions that helped us to significantly improve the present paper. This study was funded in part by the Coordena\c c\~ao de
Aperfei\c coamento de Pessoal de N\'ivel Superior - Brasil (CAPES) -
Finance Code 001, Conselho Nacional de Desenvolvimento Cient\'ifico e Tecnol\'ogico (CNPq) and Funda\c c\~ao de Amparo \`a pesquisa do Estado do RS (FAPERGS). ADM acknowledges financial support from the Spanish MCIU grant PID2019-106027GB-C41 and from the State Agency for Research of the Spanish MCIU through the ``Center of Excellence Severo Ochoa'' award for the Instituto de Astrofísica de Andalucía (SEV-2017-0709). ADM also acknowledges the support of the INPhINIT fellowship form ``la Caixa'' Foundation (ID 100010434), under the fellowship code LCF/BQ/DI19/11730018. SDSS is managed by the Astrophysical Research Consortium for the Participating Institutions of the SDSS Collaboration including the Brazilian Participation Group, the Carnegie Institution for Science, Carnegie Mellon University, the Chilean Participation Group, the French Participation Group, Harvard-Smithsonian Center for Astrophysics, Instituto de Astrofisica de Canarias, The Johns Hopkins University, Kavli Institute for the Physics and Mathematics of the Universe (IPMU) / University of Tokyo, the Korean Participation Group, Lawrence Berkeley National Laboratory, Leibniz Institut f\"ur Astrophysik Potsdam (AIP), Max-Planck-Institut f\"ur Astronomie (MPIA Heidelberg), Max-Planck-Institut f\"ur Astrophysik (MPA Garching), Max-Planck-Institut f\"ur Extraterrestrische Physik (MPE), National Astronomical Observatories of China, New Mexico State University, New York University, University of Notre Dame, Observat\' orio Nacional / MCTI, The Ohio State University, Pennsylvania State University, Shanghai Astronomical Observatory, United Kingdom Participation Group, Universidad Nacional Aut\'onoma de M\'exico, University of Arizona, University of Colorado Boulder, University of Oxford, University of Portsmouth, University of Utah, University of Virginia, University of Washington, University of Wisconsin, Vanderbilt University, and Yale University.
\end{acknowledgements}

\bibliographystyle{aa}
\bibliography{adm_r1} 

\end{document}